\documentclass[aps, a4paper,superscriptaddress, groupedaddress, floatfix,nofootinbib, twocolumn, 10pt]{revtex4}
\usepackage{amsmath,amssymb,bm,natbib}
\usepackage{epsfig}
\usepackage{graphicx}
\usepackage{slashed}

 \usepackage{color,colordvi}
 \usepackage{subfigure}
 \usepackage{float}
 \usepackage{booktabs,multirow,array}
 \usepackage{multirow}
 \usepackage{slashed}

\renewcommand{\{}{\left\lbrace}
\renewcommand{\}}{\right\rbrace}

\newcommand{\nn}{\nonumber}

\begin{document}

\phantom{}

\title{Optimal renormalization and the extraction of the strange quark mass from moments of the $\tau$-decay spectral function}

\author{B. Ananthanarayan}
\affiliation{Centre for High Energy Physics,
Indian Institute of Science, Bangalore 560 012, India}
\author{ Diganta Das}
\affiliation{Physical Research Laboratory,  Navrangpura, Ahmedabad 380 009, India}

\begin{abstract}
We introduce an optimal renormalization group analysis pertinent to the analysis of polarization 
functions associated with the $s$-quark mass relevant in $\tau$-decay. The technique is based on the 
renormalization group invariance constraints which lead to closed form summation of all the leading and next-to-leading 
logarithms at each order in perturbation theory.
The new perturbation series exhibits reduced sensitivity to the renormalization scale and improved behavior in the 
complex plane along the integration contour. Using improved experimental and theory inputs, we have extracted the value of the strange quark mass
$m_s(2{\rm GeV}) = 106.70 \pm 9.36~{\rm MeV}$ and $m_s(2{\rm GeV}) = 74.47 \pm 7.77~{\rm MeV}$
from presently available ALEPH and OPAL data respectively. These determinations are in agreement with the determinations in other 
phenomenological methods and from the lattice.

 \end{abstract}

\maketitle

\section{Introduction}
The mass of the $s$-quark is one of the fundamental parameters of the standard model \cite{Agashe:2014kda}.
It has recently been concluded that the most reliable determination comes from lattice
simulations; for a review see, e.g., \cite{Aoki:2016frl}.  However, it may be noted that the lattice determination lies
significantly below the central value given by phenomenology.  Thus, it is necessary
to use as many windows of opportunity as possible in order to make a phenomenological
extraction and to compare it with lattice results.  

Some years ago, the availability of polarization functions relevant to this extraction
was measured by the ALEPH \cite{Barate:1997hv, Schael:2005am} and OPAL \cite{Abbiendi:2004xa} Collaborations.  
The experimental collaborations themselves \cite{Barate:1997hv,Barate:1999hj} have used these measurements
to extract the $s$-quark mass. Our work is partly motivated by Ref.~\cite{Korner:2000wd}, while other theoretical approaches have been
advocated in \cite{ Pich:1998yn, Chetyrkin:1998ej, Baikov:2004tk, Chen:2001qf, Kambor:2000dj, 
Gamiz:2002nu, Gamiz:2004ar, Pich:1999bs, Davier:2001jy}.
The persistent issue in such determinations is to account 
for the renormalization group(RG) running of all the parameters that enter
the evaluations. Authors employ various methods, in order
to account for these effects when a suitable contour integral is performed in the
complex energy squared plane with chosen weight functions that define moments
of the polarization function. In \cite{Barate:1999hj,Pich:1998yn,Kambor:2000dj,Gamiz:2002nu} on the other hand, 
the QCD expressions were replaced by the phenomenological parametrizations.
In all pieces of work, suitable moments are considered which are 
then made to yield a value of the strange quark mass. However, there is a significant spread among 
the results, and one is left wondering how stable such determinations are.  

In the context of the extraction of $\alpha_s$ from $\tau$-lepton decays, there have been
two contesting schools for treating the running, which are the so-called \emph{fixed-order
perturbation theory} (FOPT) \cite{Jamin:2005ip,Beneke:2008ad} and \emph{contour-improved perturbation theory} (CIPT) 
\cite{Pivovarov:1991bj, LeDiberder:1992jjr,LeDiberder:1992zhd},
which lead to determinations of $\alpha_s$ which do not quite agree with one another,
and the theoretical error becomes dominated by this. Within CIPT itself there is another renormalization scheme where the 
perturbation corrections to a given series are absorbed by defining 
effective parameters such as effective coupling and effective mass. This is called the \emph{method of effective charges} (MEC) 
and was used in Ref.~\cite{Groote:1997kh} to improve renormalization scale behavior of $\tau$-decay polarization functions. 

Quite generally another approach that allows one to account for the RG running is to appeal to summation
of all RG-accessible logarithms which is often referred to as optimal renormalization. 
The framework was originally proposed in \cite{Maxwell:1999dv,Maxwell:2000mm,Maxwell:2001he} and 
further generalized in Refs.~\cite{Ahmady:2002fd,Ahmady:2002pa}. The RG-accessible
logarithms are defined as the leading and next-to-leading logarithms at each order in perturbation
theory that can be accessed through the RG equation. The RG summation captures the effects that are numerically encoded
in the CIPT approach, and remains an analytical method, and does not require
numerical integration. The ingredients that are required here are the $\beta$-function coefficients as well as the expansion 
coefficients of the mass Adler function in QCD which are now known to higher orders.
The method of RG summation has been applied to other processes where 
masses of quarks do enter, such as semileptonic decays of $B$-mesons.
It has been previously used to extract the strong coupling constant $\alpha_s$ from $\tau$-decay in 
Refs.~\cite{Abbas:2012py,Abbas:2012fi,Abbas:2013eba,Abbas:2013usa} and
in Ref.~\cite{Ahmed:2015sna} it was applied to reduce renormalization scale dependence due to higher order QCD corrections
in Higgs boson production.

One of the important objectives of the present work is to resort to such a \emph{renormalization group summed perturbation theory} (RGSPT) for
quantities relevant to the extraction of strange quark mass residing in the moments of the $\tau$-decay spectral function.
We will isolate combinations that involve the $s$-quark mass and polarization
functions and apply RG invariance constraints which will
give rise to a series of nested equations which can be solved with appropriate boundary conditions in closed form
at the desired order in perturbation theory. We note here that the contribution of the condensate has not changed 
much since the work of Ref.~\cite{Pich:1998yn} and we have used the same input for our extraction of $m_s$.

With these ideas in mind, we now proceed to evaluate the $s$-quark mass $m_s$ using
the method of summation of leading logarithms based on the optimal renormalization group.
The paper is organized as follows. In Sec.~\ref{sec:form} we describe the formalism of Cabbibo suppressed 
semileptonic $\tau$-decay. In Sec.~\ref{sec:closed} we give the derivations of the closed form summations of leading and next-to-leading
logarithms of the $\tau$-decay polarization functions. The extraction of strange quark mass
from ALEPH and OPAL data is discussed in Sec.~\ref{sec:numerical} and we conclude in Sec.~\ref{sec:conclusions}.
Important formulas and expressions are collected in Appendices~\ref{app:betagamma}, \ref{app:spectral}, and \ref{app:closed}. 

\section{Formalism \label{sec:form}}
The object that describes the dynamics of semileptonic $\tau$-decay is the correlator of the two hadronic currents
\begin{eqnarray}
 \Pi_{\mu\nu}(q^2) &=& i\displaystyle\int~dx e^{iqx}\langle T j_\mu(x) j_\nu^\dagger(0) \rangle \, ,
\end{eqnarray}
where the hadronic $V$-$A$ current is $j_\mu(x)=\bar{u}\gamma_\mu(1-\gamma_5)s$ and $N_c=3$ are the numbers of colors in QCD. 
The correlator has the following Lorentz decomposition:
\begin{eqnarray}
 \Pi_{\mu\nu}(q^2) &=& \frac{N_c}{6\pi^2}\Big(q_\mu q_\nu \Pi^q(q^2) + g_{\mu\nu}\Pi^g(q^2)\Big)\, .
\end{eqnarray}
The functions $\Pi^{g}(q^2)$ and $\Pi^{q}(q^2)$ are related to the scalar functions $\Pi^{T}(q^2)$ and $\Pi^{L}(q^2)$ 
that correspond to spin 1 and spin 0 contributions of the final states hadrons
\begin{equation}
 \Pi^{T} = -\frac{\Pi^{g}}{q^2}\, ,\quad \Pi^{L} = \Pi^{q} + \frac{\Pi^{g}}{q^2}\, .
\end{equation}
The polarization functions $\Pi^{g}(q^2)$ and $\Pi^{q}(q^2)$ are computable in the perturbation theory
and receive corrections due to light quark masses. Assuming the up and the down type quarks to be very light, the dominant contribution
comes from the strange quark mass $m_s$. Keeping only the leading term in powers of $m_s^2/q^2$, the two functions can be written as
\begin{eqnarray}
 \Pi^q(q^2)&=&\Pi^q_{0}(q^2)+3\frac{m_s^2}{q^2}\Pi^{mq}(q^2)\, ,\nn \\
 \Pi^{g}(q^2)&=&-q^2\Pi^g_{0}(q^2)+\frac{3}{2} m_s^2\Pi^{mg}(q^2)\, .
\end{eqnarray}
Here $\Pi^{q,g}_0(q^2)$ are the invariant functions in the massless limit and $\Pi^{mq,mg}(q^2)$ are the quadratic strange quark mass 
correction terms. The current conservation implies $\Pi^g_0 = \Pi^q_0=\Pi_0$. In the large $Q^2$ region, where $Q^2=-q^2$, the polarization 
functions are computable in the operator product expansion (OPE) framework in terms of a series of local gauge invariant operators of 
increasing dimension $D=2n$, multiplied by the inverse power of $Q^2$. In the OPE framework, $\Pi_0$ has been calculated to $\alpha_s^3$ 
in Refs.~\cite{Gorishnii:1990vf,Surguladze:1990tg,Chetyrkin:1996ez}
and the function $\Pi^{mg}(q^2)$ is known to order $\alpha_s^2$ \cite{Gorishnii:1986pz,Surguladze:1994bx,Chetyrkin:1996hm}. 
The results for $\Pi^{mq}(q^2)$ to order $\alpha_s^2$ are given in \cite{Chetyrkin:1993hi}
and the $\alpha_s^3$ coefficients were calculated in Ref.~\cite{Baikov:2004tk}. 

Formally, the polarization functions depend on the renormalization scale $\mu$ through 
the strong coupling constant $\alpha_s(\mu^2)$ and logarithmic terms $\ln(\mu^2/Q^2)$, in the form of $\alpha_s^n(\mu^2)\ln^{n-k}(\mu^2/Q^2)$,
at each order $n$ in the perturbation expansion. Unless explicitly mentioned, we will denote the functional dependence of 
$\mu$ as $\Pi^{mg,mq}[\mu^2,Q^2] = \Pi^{mg,mq}$. The scale evolution of any perturbation series is described by the renormalization group 
equation. It is known that the function $\Pi^{mg}$ is not renormalization group invariant and obeys the following RG equation 
\cite{Chetyrkin:1996hm}:
\begin{equation}\label{eq:Pimg:RG2}
 \frac{\partial \Pi^{mg}}{\partial L} + \beta \frac{\partial \Pi^{mg}}{\partial a_s} + 2 \gamma_m \Pi^{mg} = \frac{\gamma_m^{VV}}{6} \, ,
\end{equation}
where $L = \ln(\mu^2/Q^2)$, $a_s=\alpha_s(\mu^2)/\pi$, and the anomalous dimension matrix $\gamma_m^{VV}$ is given 
in \cite{Chetyrkin:1996hm,Chetyrkin:1996sr}.
In the renormalization group equations above, $\beta(a_s)$ and $\gamma_m(a_s)$ encode the scale evolution of the strong coupling constant 
and mass, respectively, and the evolution equations are given as 
\begin{eqnarray}\label{def:beta}
 \mu^2\frac{da_s(\mu^2)}{d\mu^2} &=& \beta(a_s) = -\sum_{i=0} \beta_i a_s^{i+2} \, ,\\
 \label{def:gamma}
 \mu^2 \frac{dm(\mu^2)}{d\mu^2} &=& m \gamma_m(a_s) = -m\sum_{i=0}\gamma_i a_s^{i+1}\, ,
\end{eqnarray}
where
\begin{equation}
 \mu^2 \frac{d}{d\mu^2} = \frac{\partial}{\partial L} + \beta(a_s) \frac{\partial}{\partial a_s} + 
 \gamma_m(a_s) m_s \frac{\partial}{\partial m_s} \, .\nn\\
\end{equation}
The evolution equations can be solved iteratively in terms of the coefficients $\beta_i$ and $\gamma_i$ and the solutions
are given in Appendix~\ref{app:betagamma}. Corresponding to the function $\Pi^{mg}$, one can define an Adler function 
\cite{Baikov:2002uw}:
\begin{equation}\label{eq:defAdler}
 D^{mg} = D^{mg}[\mu^2,Q^2] = -Q^2\frac{d\Pi^{mg}}{dQ^2} \, .
\end{equation}
which satisfies the homogeneous RG equation
\begin{equation}\label{eq:RGD}
 \frac{\partial D^{mg}}{\partial L} + \beta \frac{\partial D^{mg}}{\partial a_s} + 2 \gamma_m D^{mg} = 0\, .
\end{equation}
The Adler function $D^{mg}$ is known to order $\alpha_s^3$ in perturbation theory
and the $\alpha_s^4$ coefficients are only partially known \cite{Baikov:2002uw}.
The polarization function $\Pi^{mq}$ is also satisfies the homogeneous RG equation
\begin{equation}\label{eq:RGPimq}
 \frac{\partial \Pi^{mq}}{\partial L} + \beta \frac{\partial \Pi^{mq}}{\partial a_s} + 2 \gamma_m \Pi^{mq} = 0\, .\\
\end{equation}
Note that the expressions of $D^{mg}$ and $\Pi^{mq}$ given in Refs.\cite{Baikov:2002uw}
and \cite{Baikov:2004tk, Korner:1999kw}, respectively, are evaluated at the normalization scale $\mu^2 = Q^2$ which makes all the 
logarithmic terms $\ln(\mu^2/Q^2)$ vanish.  To apply the resummation method to take into account of the higher order perturbation
corrections, we have generated these logarithmic terms using
the renormalization group equations (\ref{eq:RGD}) and (\ref{eq:RGPimq}). 
The expressions of $\Pi^{mg,mq}$ and $D^{mg}$ are collected in Appendix \ref{app:spectral}.


The Cabbibo suppressed $\tau$-decay results in final states that are induced by the vector and axial vectors currents. In experimental studies 
of Cabbibo suppressed $\tau$-decay, the measurements of $SU(3)$ breaking effects induced by strange quark mass have been possible by separately 
measuring the strangeness changing and strangeness conserving $\tau$-decay rates. In addition to the total decay rate, important quantities are 
the moments which can be written as a
contour integral of the polarization functions with suitable weight functions in the complex  $|q^2|$ plane
running counterclockwise along the circle $|q^2|=M_\tau^2$ \cite{Braaten:1991qm, Braaten:1988ea, Narison:1988ni}
\begin{align}\label{eq:Rkl}
& R^{kl}_{\tau} = \frac{6i}{2\pi} \oint \rho_{k,l}(q^2) \Bigg[ \frac{m_s^2}{q^2}\Pi^{mq} - \frac{m_s^2}{M_\tau^2}\Pi^{mg} \Bigg] \frac{dq^2}{M_\tau^2}\, .
\end{align}
Here $m_s = m_s(\mu^2)$ and $\rho_{k,l} = (1-q^2/M_\tau^2)^{k+2}(q^2/M_\tau^2)^l$ are suitable weight functions. The moments equation (\ref{eq:Rkl}) 
are physical quantities and therefore independent of the renormalization scale. 
However, one of the main sources of theoretical uncertainty in the determinations of the moments come from the renormalization scale 
dependence of the perturbation series. The scale dependence arises due to finite terms in the perturbation series.
This uncertainty can be partly reduced by renormalization group improvements of the perturbation series
through the method of resummation of logarithmic terms. 

As previously mentioned, there are several competing methods to perform the renormalization group improvements which lead to slightly different results. 
In the FOPT, it amounts to first performing the contour integration in Eq.~(\ref{eq:Rkl}), 
followed by renormalization group improvement by setting the renormalization scale $\mu^2 = M_\tau^2$ \cite{Jamin:2005ip,Beneke:2008ad}.
In CIPT, the renormalization group 
improvement is achieved by setting $\mu^2 = Q^2$ followed by the integration along the contour. The fixing of the renormalization scale at $\mu^2=Q^2$
before the integration essentially means that at each order in the perturbation expansion, the series depends on a constant term. To account for this, 
in CIPT the strong coupling constant and the mass are evolved along the contour in the complex plane. This is done by iteratively solving the evolution 
equations along the integration contour with initial conditions $\alpha_s(M_\tau^2)$ and $m_s(M_\tau^2)$ at $Q^2 = M_\tau^2$.

The resummation of the polarization functions to the quadratic mass corrections were treated in MEC in 
Ref. \cite{Korner:2000wd} following a partial integration in Eq. (\ref{eq:Rkl}) 
\begin{eqnarray}\label{eq:Rkl2}
R^{kl}_{\tau} &=&  \frac{6i}{2\pi} \oint \rho_{k,l}(q^2) \frac{m^2_s}{q^2} \Pi^{mq} \frac{dq^2}{M_\tau^2}\, \nn\\ 
 &-& \frac{6i}{2\pi} \oint \Bigg[ \frac{m_s^2}{q^2}D^{mg} \int \rho_{k,l}(q^2) \frac{dq^2}{M_\tau^2}  \Big] \frac{dq^2}{M_\tau^2} \, .
\end{eqnarray}
In Ref.~\cite{Korner:2000wd} the higher order perturbative corrections to the polarization function $\Pi^{mq}$ and the Adler function $D^{mg}$
were absorbed in the definition of the effective strong coupling constant and two coefficients of effective mass parameters.
Note that, renormalization group improvements and partial integration do not commute as long the perturbation series in question is known
only to finite order. Since the work in \cite{Korner:2000wd}, the $\alpha_s^3$ coefficient of the polarization function has been calculated in \cite{Baikov:2004tk}. Using these
new coefficients and the latest inputs, we update the strange quark mass in MEC to order $\alpha_s^3$.

However, the main aim of the work is to apply the optimal renormalization group to sum up RG-accessible
logarithms of the $\tau$-decay polarization function. To the best of our knowledge, the optimal renormalization group is being 
applied to $\tau$-decay in the context of strange quark mass extraction for the first time. 
In the next section, we derive the closed form
summation of the $\tau$-decay polarization functions and show that these RGSPT series exhibit reduced RG scale dependence. The new series is then 
used in Sec.~\ref{sec:numerical} to extract strange quark mass from ALEPH and OPAL data.

\section{Closed form sum of RG-accessible logarithms \label{sec:closed}}
Here we develop RGSPT relevant to the extraction of the $s$-quark mass.
To apply the framework we write the perturbation series in question as
\begin{eqnarray}\label{eq:PimqPert}
\Pi^{mg} &=& \sum_{n=0}^\infty \sum_{k=0}^n c^{mg}_{n,k} a_s^n L^k \, ,\nn\\
D^{mg} &=& \sum_{n=0}^\infty \sum_{k=0}^n d^{mg}_{n,k} a_s^n L^k \, ,\\
 \Pi^{mq} &=& \sum_{n=0}^\infty \sum_{k=0}^n c^{mq}_{n,k} a_s^n L^k \, \nn ,
\end{eqnarray}
where $c^{mg,mq}, d^{mg}$ are the series coefficients. The coefficients can be easily determined from the expressions
given in Appendix~\ref{app:spectral}. 
Following Ref.~\cite{Ahmady:2002fd} the RGSPT series are written as 
\begin{eqnarray}\label{eq:RGSI}
\Pi^{mg}_{{\rm RGSPT}} &=& \sum_{n=0}^{\infty} a_s^n \Pi^{mg}_n[a_s L]\, ,\nn\\
D^{mg}_{{\rm RGSPT}} &=& \sum_{n=0}^{\infty} a_s^n D^{mg}_n[a_s L]\, ,\\
 \Pi^{mq}_{{\rm RGSPT}} &=& \sum_{n=0}^{\infty} a_s^n \Pi^{mq}_n[a_s L]\, ,\nn
\end{eqnarray}
where the intermediate quantities $\Pi^{mg,mq}_k[a_s L]$ and $D_k^{mg}[a_s L]$ are defined as
\begin{equation}\label{eq:PiD:2}
\begin{split}
\Pi_k^{mg}[a_s L] = \sum_{n=k}^\infty c^{mg}_{n,n-k}(a_s L)^{n-k}\, ,\\
D_k^{mg}[a_s L] = \sum_{n=k}^\infty d^{mg}_{n,n-k}(a_s L)^{n-k}\, ,\\
 \Pi_k^{mq}[a_s L] = \sum_{n=k}^\infty c^{mq}_{n,n-k}(a_s L)^{n-k}\, .\\
\end{split}
\end{equation}
The RG equations (\ref{eq:Pimg:RG2}), (\ref{eq:RGD}), and (\ref{eq:RGPimq}) ensure closed form expressions of 
$\Pi_k^{mg}[a_s L]$, $D_k^{mg}[a_s L]$, and $\Pi^{mq}_k[a_s L]$, respectively, in terms of the leading coefficients $d^{mg}_{n,0}$ and $c^{mg,mq}_{n,0}$. 
To illustrate how it works, we substitute $\Pi^{mg}$ [Eq.~(\ref{eq:PimqPert})]
in the RG equation (\ref{eq:Pimg:RG2}) and collect the coefficients of $a_s^n L^{n-1-k}$, which leads to the following recursion relation:
\begin{align}\label{eq:rec}
& (n-k) c^{mg}_{n,n-k}  -\sum_{i=0}^\infty \beta_i (n-i-1) c^{mg}_{n-i-1,n-1-k} \nn\\
& - 2 \sum_{i=0}^\infty \gamma_i c^{mg}_{n-i-1,n-1-k}  =  t_{n,n-1-k}  \, . 
\end{align}
The relation is subject to the condition that for $p<q$, $c^{mg}_{p,q} = 0$. In this equation $t_{n,n-1-k}$ are the perturbation series coefficients
of the left-hand side of the equation (\ref{eq:Pimg:RG2}) which is defined as  $\gamma_m^{VV} = 6\sum_{n=0}^\infty \sum_{k=0}^n t_{n,k} a_s^n L^k$.
We multiply Eq. (\ref{eq:rec}) by $(a_sL)^{n-1-k}$ and sum from $n=1+k$ to infinity, which following Eq.~(\ref{eq:PiD:2}) results in a differential
equation for $\Pi^{mg}_k[a_s L]$. The differential equations are solved with the boundary conditions
$\Pi^{mg}_k[0] = c^{mg}_{k,0}$ and the resultant solutions are the closed form expressions of $\Pi^{mg}_k[a_s L]$. 
Since $D^{mg}$ and $\Pi^{mg}$ satisfy homogeneous RG equations, their recursion relations are obtained by putting $t_{n,n-1-k} = 0$ 
followed by the replacements $c^{mg} \to d^{mg}$ and $c^{mg} \to c^{mq}$ respectively. The rest of the derivations are similar. 
The closed form expressions of $\Pi^{mg}_{0,1,2}[a_s L]$ in terms of the variable $\omega = 1-\beta_0 a_s L$ are

\begin{eqnarray}\label{eq:sol0}
\Pi^{mg}_0[a_s  L] &=& \Big(c^{mg}_{0,0}+\frac{t_{1,0}}{2\gamma_0}\Big)\omega^{-A}-\frac{t_{1,0}}{2\gamma_0} \, ,\\
\Pi^{mg}_1[a_s  L] &=& \frac{T_1}{1+A}\{1- \omega^{-A-1} \} + (B+T_2) \omega^{-A} \, \\
&+&\{ c^{mg}_{1,0} - (B+T_2) + (C+T_3) \ln(\omega) \}\omega^{-A-1}\, \nn
\end{eqnarray}
\begin{widetext}
\begin{eqnarray}
\Pi^{mg}_2[a_s  L] &=& \frac{T_4}{2+A}+\frac{D+T_5}{2}\omega^{-A} + (E+T_6-F-T_7)\omega^{-1-A}+(F+T_7)\omega^{-1-A}\ln(\omega)  \nn\\ 
 &+& \Big(c^{mg}_{2,0} -\frac{D+T_5}{2} - E-T_6 +F + T_7 -\frac{T_4}{2+A}\Big)\omega^{-2-A}
 + (G+T_8)\omega^{-2-A}\ln(\omega) \nn\\
 &+& \frac{H+T_9}{2}\omega^{-2-A}\ln^2(\omega)
 \end{eqnarray}
The closed form expressions of $D^{mg}_{0,1,2}[a_s L]$ ($\Pi^{mq}_{0,1,2}[a_s L]$) are obtained by the replacements $c^{mg}\to d^{mg}$ ($c^{mg}\to c^{mq}$) 
and putting $t_{1,0} = 0$, $T_{1\cdots 9} =0$. Following a similar derivation the expression of $D^{mg}_3[a_s L]$ reads
 \begin{eqnarray}\label{eq:sol3}
 D^{mg}_3[a_s  L] &=& \frac{K}{3}\omega^{-A} + \Big(\frac{M}{2}-\frac{N}{4}\Big)\omega^{-1-A} + \frac{N}{2}\omega^{-1-A}\ln\omega + (P-Q+2R)\omega^{-2-A} + (Q-2R)\omega^{-2-A}\ln\omega \, \nn\\
 &+& R\omega^{-2-A}\ln^2\omega + \Bigg(d^{mg}_{3,0}-\frac{K}{3} - \frac{M}{2} +\frac{N}{4} -P +Q -2R \Bigg)\omega^{-3-A}
 + U\omega^{-3-A}\ln\omega + \frac{V}{2}\omega^{-3-A}\ln^2\omega \, \nn\\
 &+& \frac{Y}{3}\omega^{-3-A}\ln^3\omega
 \end{eqnarray}

\end{widetext}
The corresponding expression of $\Pi^{mq}_3[a_s L]$ is obtained by the replacement $d^{mg}_{3,0} \to c^{mq}_{3,0}$. 
In the above expressions $A=2\gamma_0/\beta_0$ and the rest of the coefficients $B$, $C$, $D$, $E$, $F$, $G$, $H$, $K$, $M$, $N$, $P$, 
$Q$, $R$, $S$, $T$, $U$, $V$, $Y$ and $T_{1\cdots 9}$ depend on the coefficients of the $\beta$ and the $\gamma_m$ functions as well as the leading 
coefficients $d^{mg}_{n,0}, c^{mg,mq}_{n,0}$.
The detailed expressions are given in Appendix~\ref{app:closed}. 
\begin{figure}[ht]
\begin{center}
 \includegraphics[width = 4.2cm]{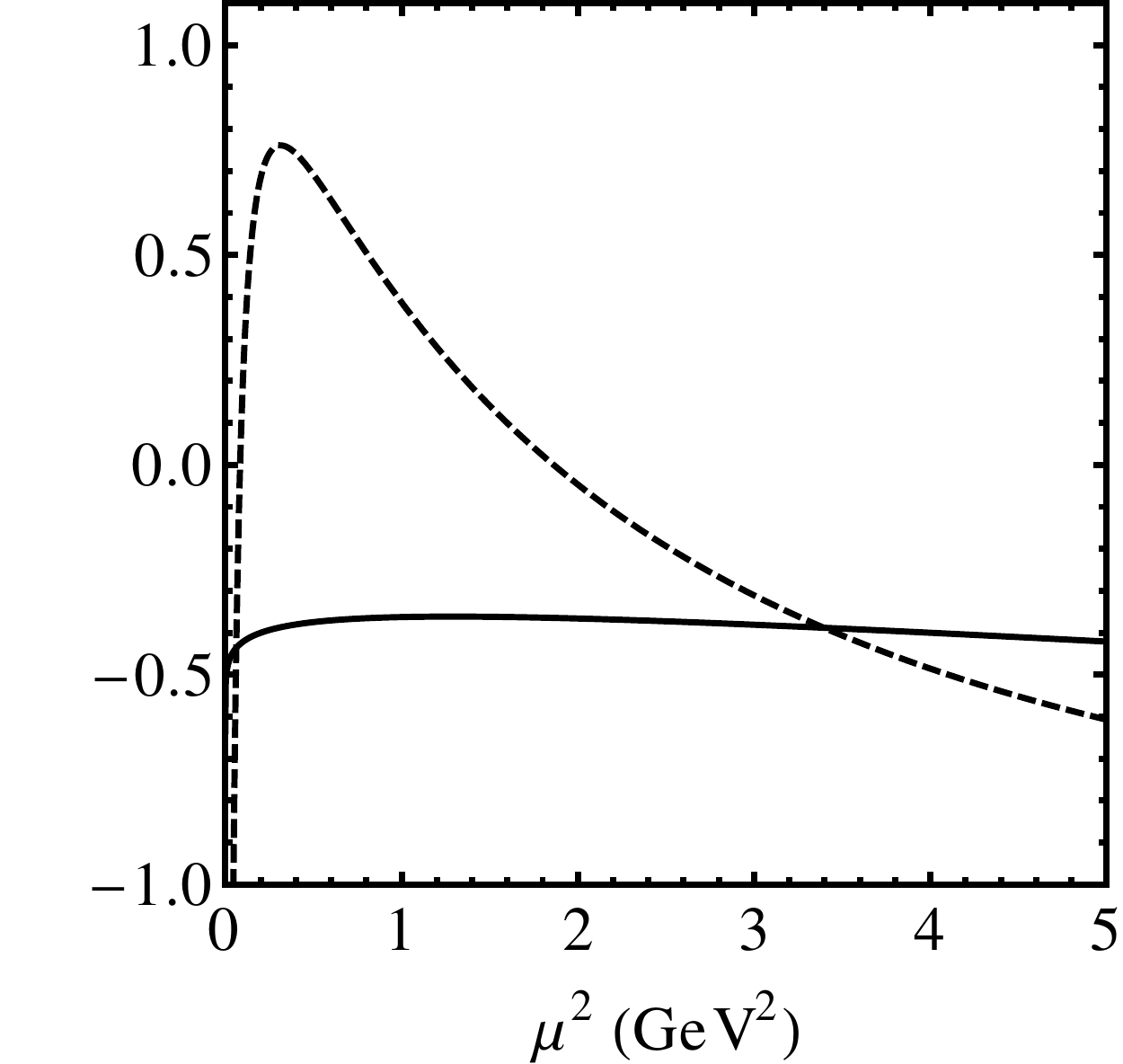}
 \includegraphics[width = 4.cm]{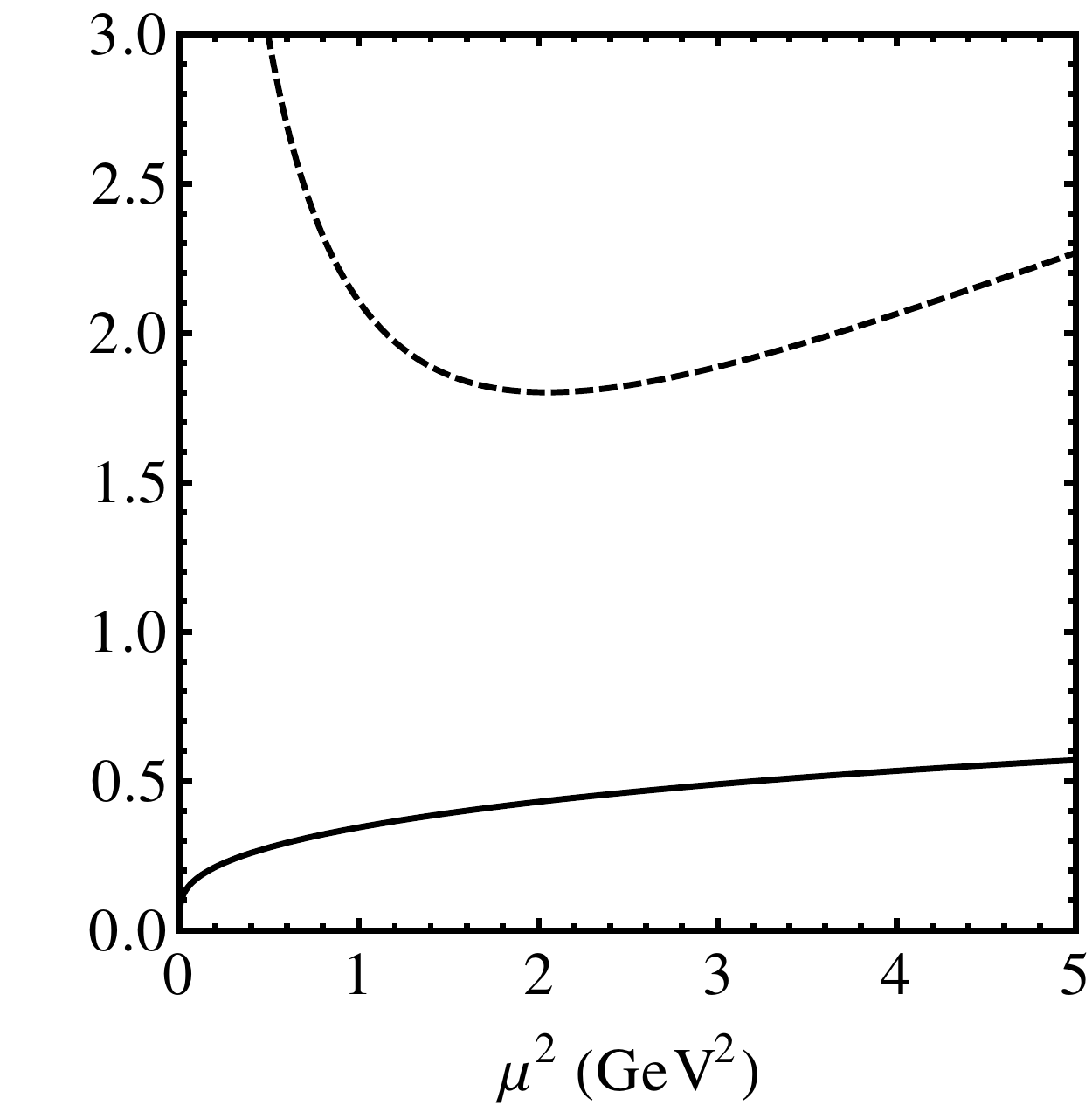}
\caption{Renormalization scale dependence of the polarization function $\Pi^{mg}[\mu^2,Q^2]$ at $\mathcal{O}(\alpha_s^2)$ for $Q^2 = -M_\tau^2$. 
The solid and the dashed lines correspond to the RG-summed equation (\ref{eq:RGSI}) and the unsummed series equation (\ref{eq:PimgUnsummed}), respectively. 
The figures to the left and right correspond to real and imaginary parts of the functions, respectively. 
\label{fig:scale:Pimgas2}}
\end{center}
\end{figure}
\begin{figure}[ht]
\begin{center}
 \includegraphics[width = 4.25cm]{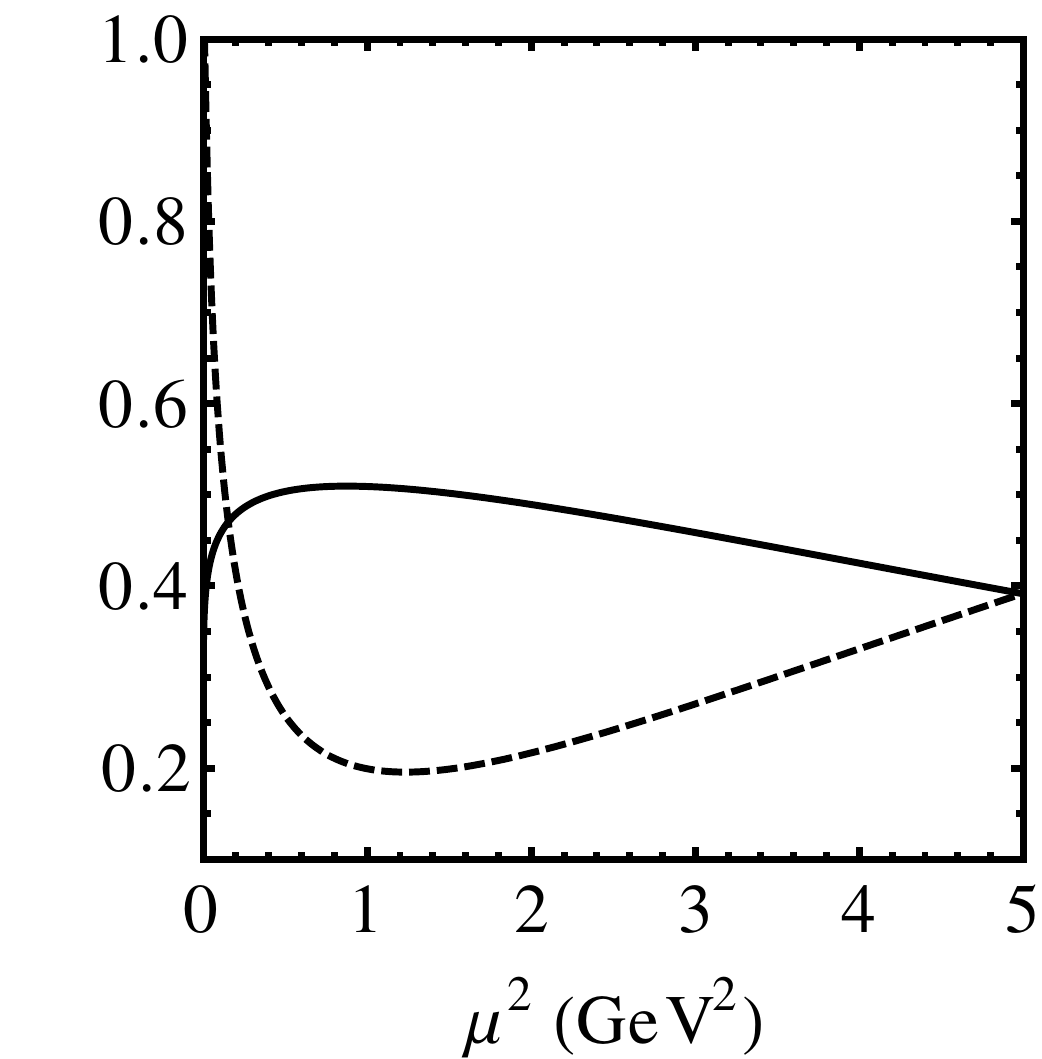}
 \includegraphics[width = 4.25cm]{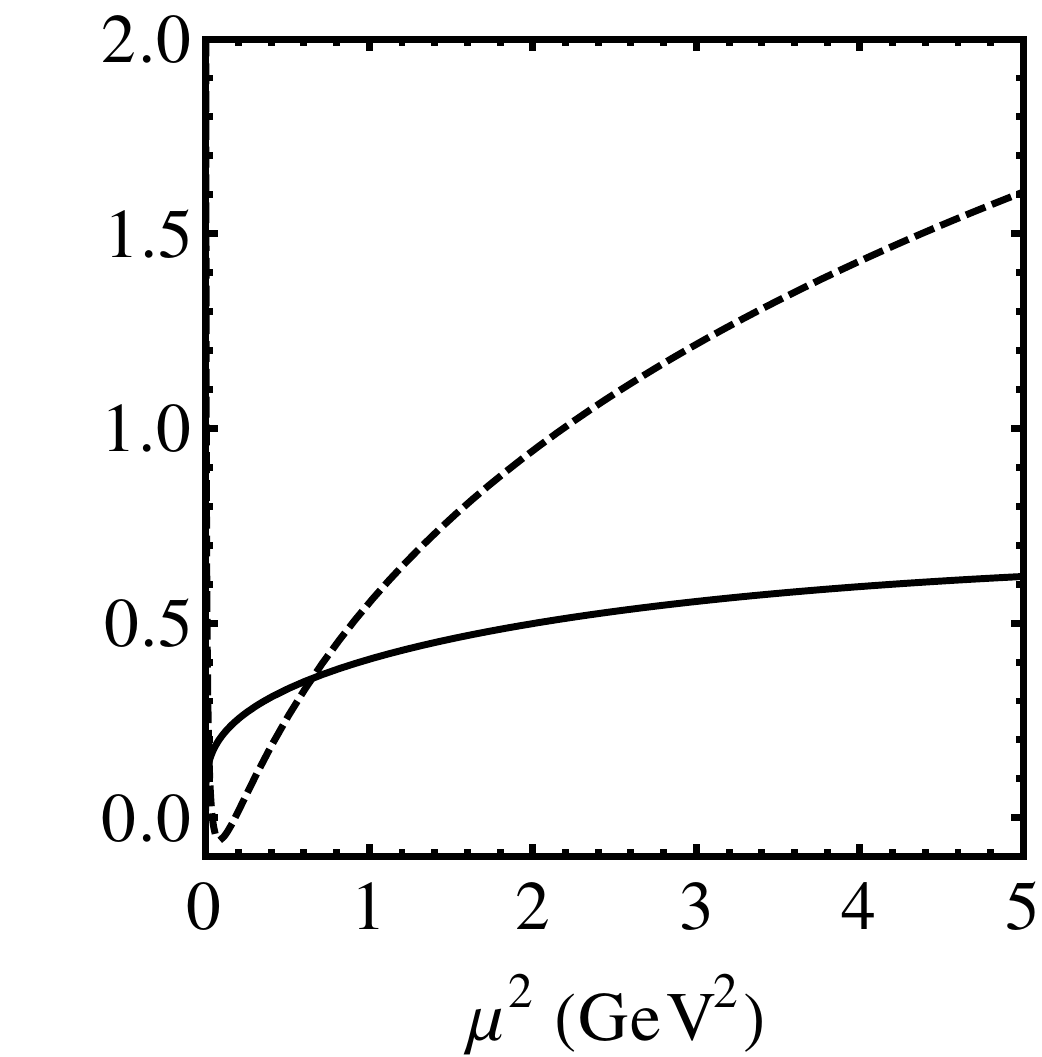}
\caption{Renormalization scale dependence of the polarization function $D^{mg}[\mu^2,Q^2]$ for $Q^2 = -M_\tau^2$. 
The solid and the dashed lines correspond to the RG-summed Eq.~(\ref{eq:RGSI}) and the unsummed series Eq.~(\ref{eq:DmgUnsummed}), respectively. 
The figures to the left and right correspond to real and imaginary parts of the functions, respectively. 
\label{fig:scale:Dmgas3}}
\end{center}
\end{figure}
\begin{figure}[ht]
\begin{center}
 \includegraphics[width = 4cm]{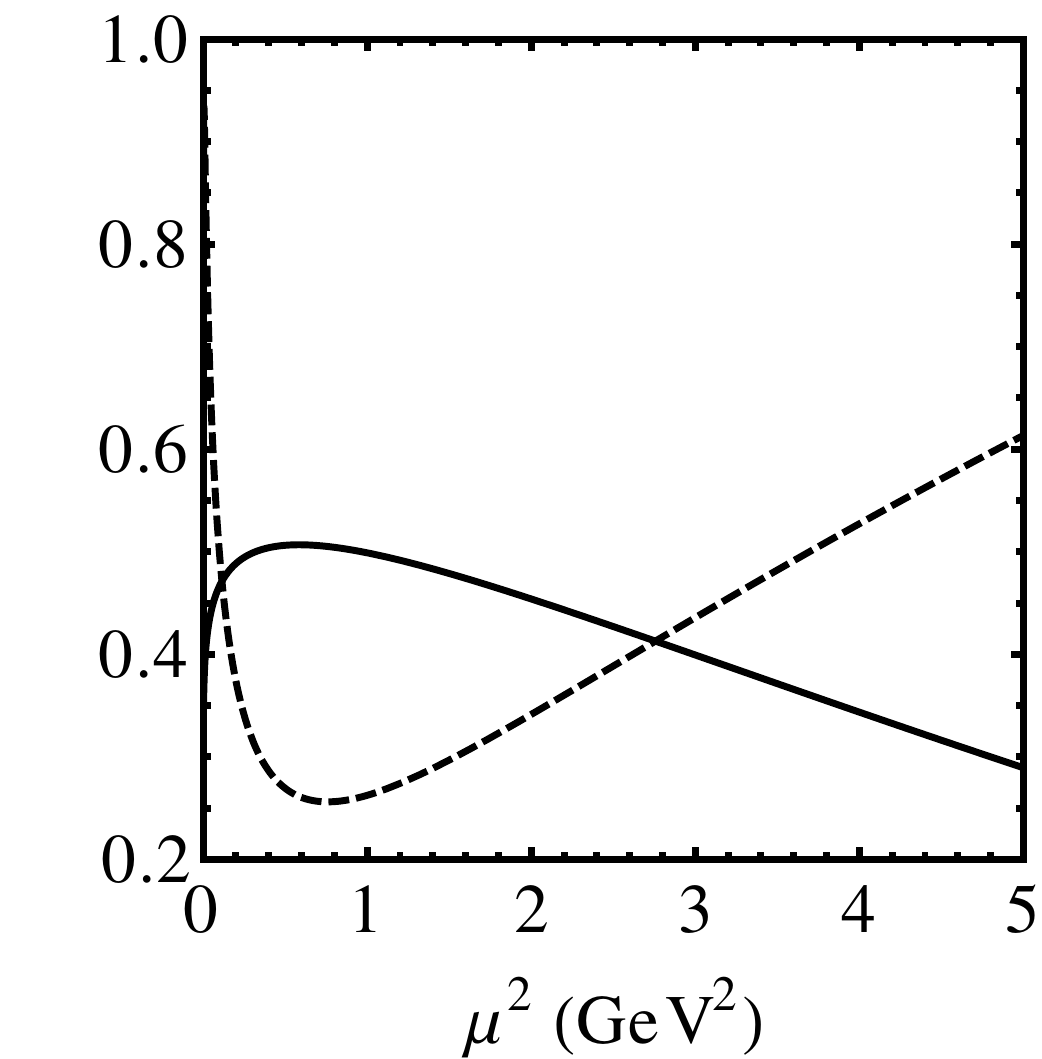}
 \includegraphics[width = 4cm]{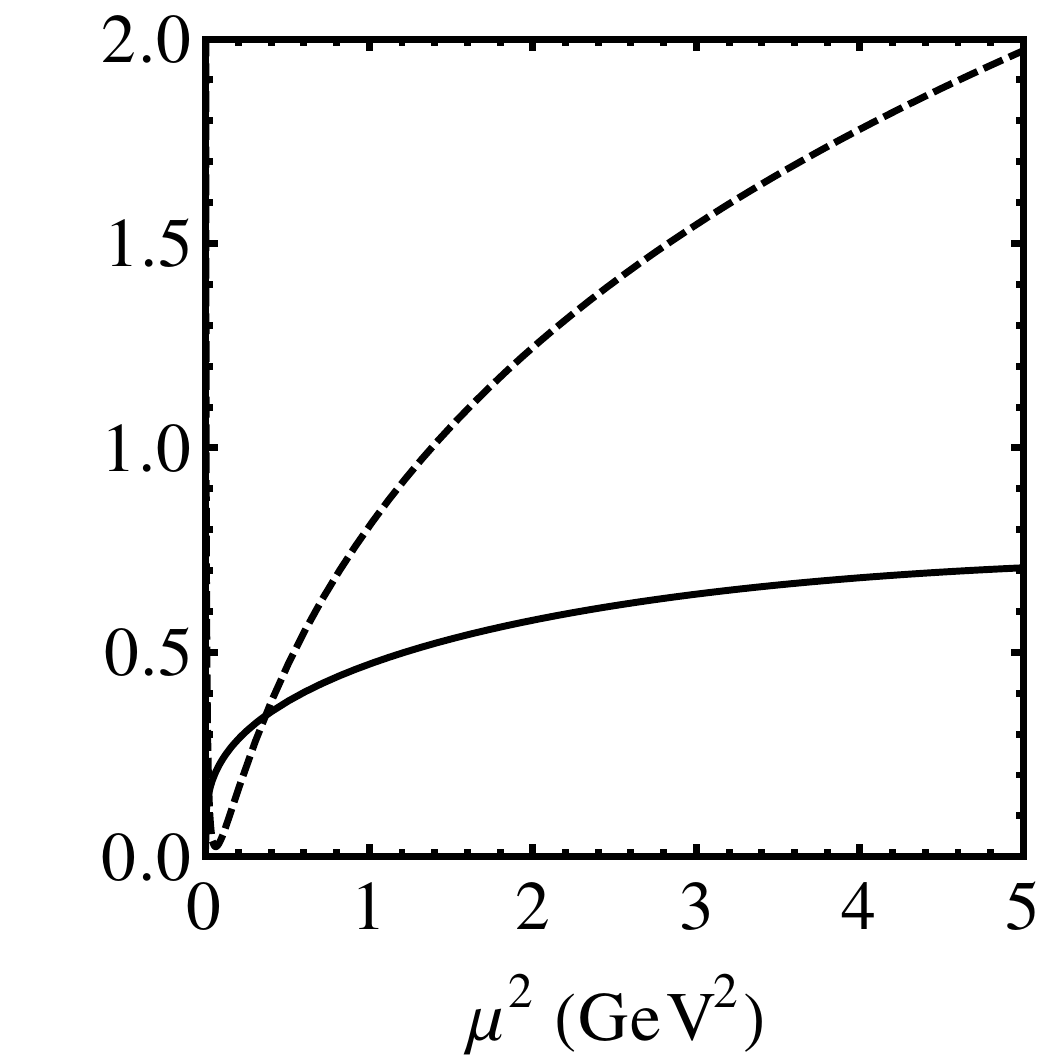}
\caption{Renormalization scale dependence of the function $\Pi^{mq}[\mu^2,Q^2]$ for $Q^2 = -M_\tau^2$. 
The solid and the dashed lines correspond to RG-summed equation (\ref{eq:RGSI}) and the un-summed series equation (\ref{eq:PimqUnsummed}), respectively. 
The figures to the left and right corresponds to real and imaginary parts of the functions, respectively. 
\label{fig:scale:Pimqas3}}
\end{center}
\end{figure}

In Fig.~\ref{fig:scale:Pimgas2} we study the scale dependence of the new series where we plot $\Pi^{mg}$ in RGSPT 
and compare with the unsummed perturbation series equation (\ref{eq:PimgUnsummed}). The RGSPT and the un-summed series are shown by solid and dashed lines,
respectively, and the figure to the left and right correspond to the real and imaginary parts of the functions. In Figs.~\ref{fig:scale:Dmgas3} 
and \ref{fig:scale:Pimqas3} we plot $D^{mg}$ and $\Pi^{mq}$ in RGSPT and compare with the corresponding 
unsummed series. In these figures we have taken $Q^2=-M_\tau^2$ and
the strong coupling constant is taken from direct measurements at the $Z$ peak, $\alpha_s(M_Z^2)=0.1197\pm 0.0028$ 
\cite{Agashe:2014kda}. This is evolved using a four-loop $\beta$ function using the package \cite{Chetyrkin:2000yt,Schmidt:2012az} to the $\tau$ mass scale which yields 
$\alpha_s(M_\tau^2)=0.3298\pm 0.0118$, and is in excellent agreement with the latest determination from $\tau$-decay in \cite{Pich:2016bdg}.
The figures clearly show that the renormalization group improved series based on the optimal 
renormalization group has substantially less scale dependence 
than the unsummed series. This reduces the uncertainty coming from the renormalization scale. 
With the formalism developed and with the expressions on hand, we now
proceed to the extraction of the strange quark mass in the next section.  

\section{Extraction of strange quark mass \label{sec:numerical}}
After obtaining the RG-summed perturbation series in Eq.~(\ref{eq:RGSI}) we can perform the integrations in Eqs.~(\ref{eq:Rkl}) and 
(\ref{eq:Rkl2}) in the complex plane along the contour defined by $Q^2 = M_\tau^2 e^{I\phi}$. In RGSPT based on the optimal renormalization 
group we set the renormalization scale
$\mu^2=M_\tau^2$ so that the strong coupling constant and the strange quark mass are evaluated at $M_\tau^2$: $\alpha_s(M_\tau^2)$,
$m_s(M_\tau^2)$. This is in contrast to the CIPT where both of these quantities are evolved along the contour, while in MEC the effective 
couplings and the coefficients of effective masses take into account the scale evolution.

In this section we extract the strange quark mass from the data on the moments of the $\tau$-decay spectral function $R^{kl}_\tau$.
Such extractions are possible from the strange quark mass induced $SU(3)$ flavor breaking term
\begin{equation}
 \delta R_\tau^{kl} = \frac{R^{kl}_{\tau,\text{non-}S}}{|V_{ud}|^2} - \frac{R^{kl}_{\tau,S}}{|V_{us}|^2}\, ,
\end{equation}
where the strange and the nonstrange contributions are denoted by $R^{kl}_{\tau,S}$ and $R^{kl}_{\tau,\text{non-}S}$, respectively. 
Theoretically, $\delta R^{kl}_\tau$ is related to the $D\ge  2$ correction terms in OPE. 
In addition to the perturbative corrections, $\delta R_\tau^{kl}$ also receive nonperturbative contributions from $D=4$ condensates 
corrections to the $\tau$-decay rate \cite{Braaten:1991qm}.
Neglecting terms of the order $m_s^3/M_\tau^3$ and higher, the theoretical expression of $\delta R^{kl}_\tau$ is given as
\begin{equation}\label{eq:ms}
 \delta R^{kl}_\tau = N_c S_{EW} \Big(-R^{kl} -4\pi^2 \frac{m_s^2(M_\tau^2)}{M_\tau} \frac{\langle\bar{s}s\rangle}{M_\tau^3}f_{kl} \Big) \, .
\end{equation}
where $S_{EW} = 1.0201 \pm 0.0003$ \cite{Erler:2002mv, Braaten:1990ef, Marciano:1988vm}. The coefficients of $D=4$ corrections to the local
$m_s\bar{s}s$ operators $f_{kl}$ in the leading order in OPE are given by \cite{Pich:1998yn}
\begin{equation}
 f_{kl} = 2(\delta_{l,0}(k+2)-\delta_{l,1})\, ,\nn
\end{equation}
and the phenomenological value for the quark condensates is \cite{Pich:1998yn}
\begin{equation}
 \langle \bar{s}s\rangle = (0.8 \pm 0.2) \langle \bar{u}u\rangle\, ,\quad \langle \bar{u}u\rangle\ = (-0.23\text{GeV})^3\, .
\end{equation}

The flavor breaking terms $\delta R^{kl}_\tau$ for different moments $(k,l)$ have been measured by ALEPH \cite{Barate:1997hv,Schael:2005am} and OPAL 
\cite{Abbiendi:2004xa} Collaborations by separately measuring
the Cabbibo-allowed and the Cabbibo-suppressed inclusive $\tau$-decay rates. For ALEPH data we follow the moments extracted in Ref.~\cite{Chen:2001qf}.
The moments with large $k$ and $l=0$ are dominated by low lying resonances which can be precisely measured in
the experiments. Theoretically however, this amounts to the calculations of higher dimensional condensate terms that are 
not presently known.  Therefore from the point of view of theoretical calculations, the moments with large $k$ and $l=0$ are 
not favorable. On the other hand, the moments of the type $(0,l)$ with large $l$ are more reliable from the point of view of
perturbation theory but their experimental errors are large.
In our analysis we calculate $m_s$ for three moments (0,0), (1,0), and (2,0) measured by ALEPH in Ref.~\cite{Chen:2001qf}. 
For OPAL the moments, (2,0), (3,0), and (4,0) are considered as 
the most reliable that we have used in our extraction.

At $\mathcal{O}(\alpha_s^2)$, the moments of the total decay rate can be calculated through direct integration
of the polarization functions in Eq.~(\ref{eq:Rkl}). One of these functions $\Pi^{mg}$ is not RG invariant but satisfies
an inhomogeneous RG equation given in Eq.~(\ref{eq:Pimg:RG2}). These determinations can be compared
with that based on the partial integration equation (\ref{eq:Rkl2}) where $\Pi^{mg}$ is replaced by an Adler function
that satisfies the homogeneous RG equation (\ref{eq:RGD}). In Table~\ref{table:RklThas2} we compare the values of $\widetilde{R}^{kl}_\tau$ defined 
as $R^{kl}_\tau=m_s^2(M_\tau^2)\widetilde{R}^{kl}_\tau/M_\tau^2$, obtained in these two methods. Here the errors correspond
to the uncertainties coming from theory inputs.
From the table it is seen that $\widetilde{R}^{kl}_\tau$ from Eq.~(\ref{eq:Rkl}) gives smaller 
values compared to that obtained from Eq.~(\ref{eq:Rkl2}). In Table~\ref{table:msas2} we extract strange quark mass 
$m_s(M_\tau^2)$ at $\alpha_s^2$ from
ALEPH data using these two versions of performing the contour integration. 
Here the first error corresponds to the uncertainties coming from experimental inputs whereas the second errors correspond
to the theory uncertainties. The table shows that there are some differences in the extracted values of $m_s$ 
depending on the version of the contour integration used and the one 
based on partial integration leads to smaller values, though they are consistent with each other within the theory errors.
It is therefore desirable that a full analysis of the polarization function also be performed,
but that requires the perturbative coefficients for this quantity, which have not appeared
in the literature. Such a result may help clarify the situation as regards to which version
leads to more stable results.

\begin{figure}[ht]
\begin{center}
 \includegraphics[width = 4cm]{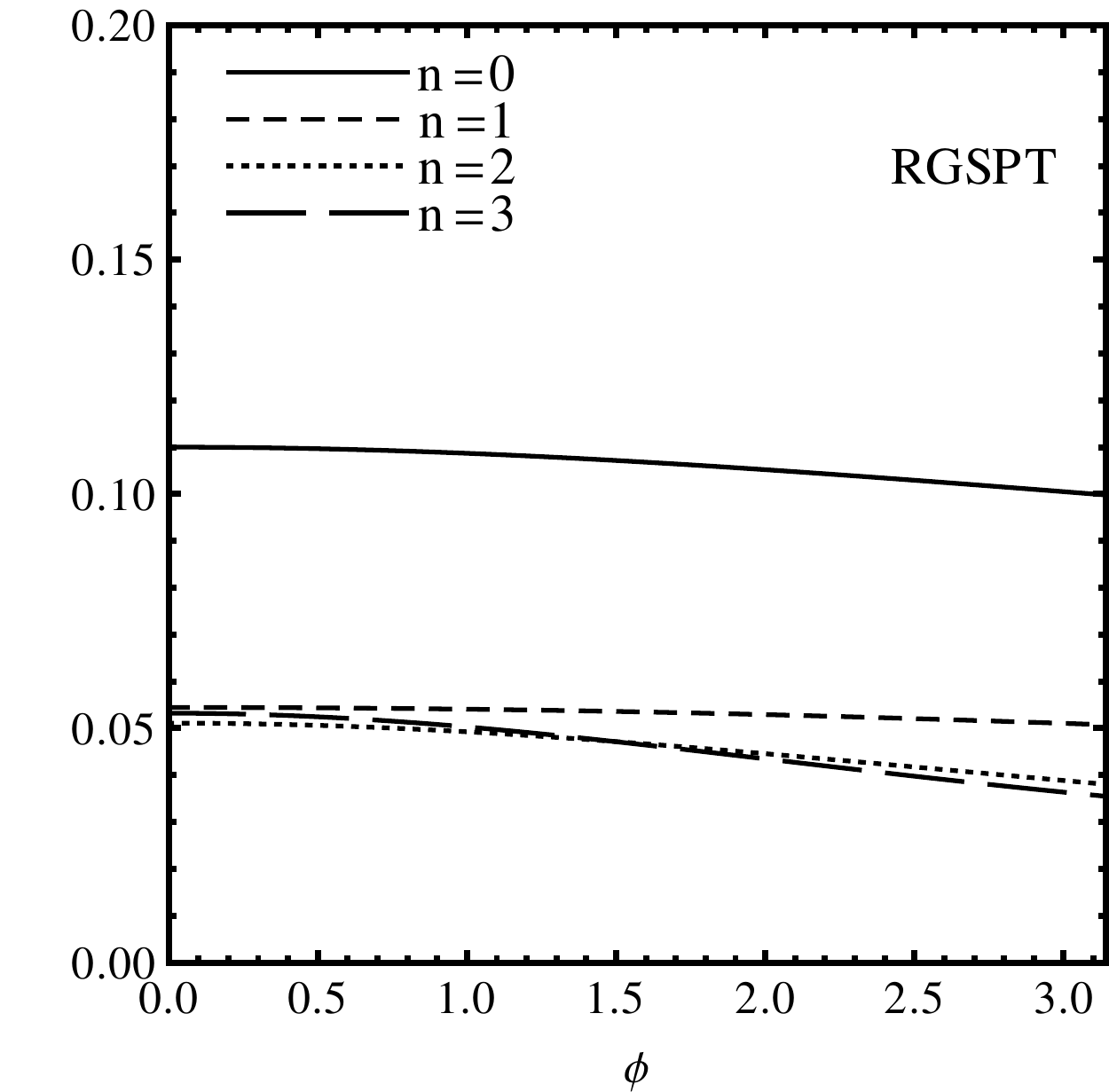}
 \includegraphics[width = 4cm]{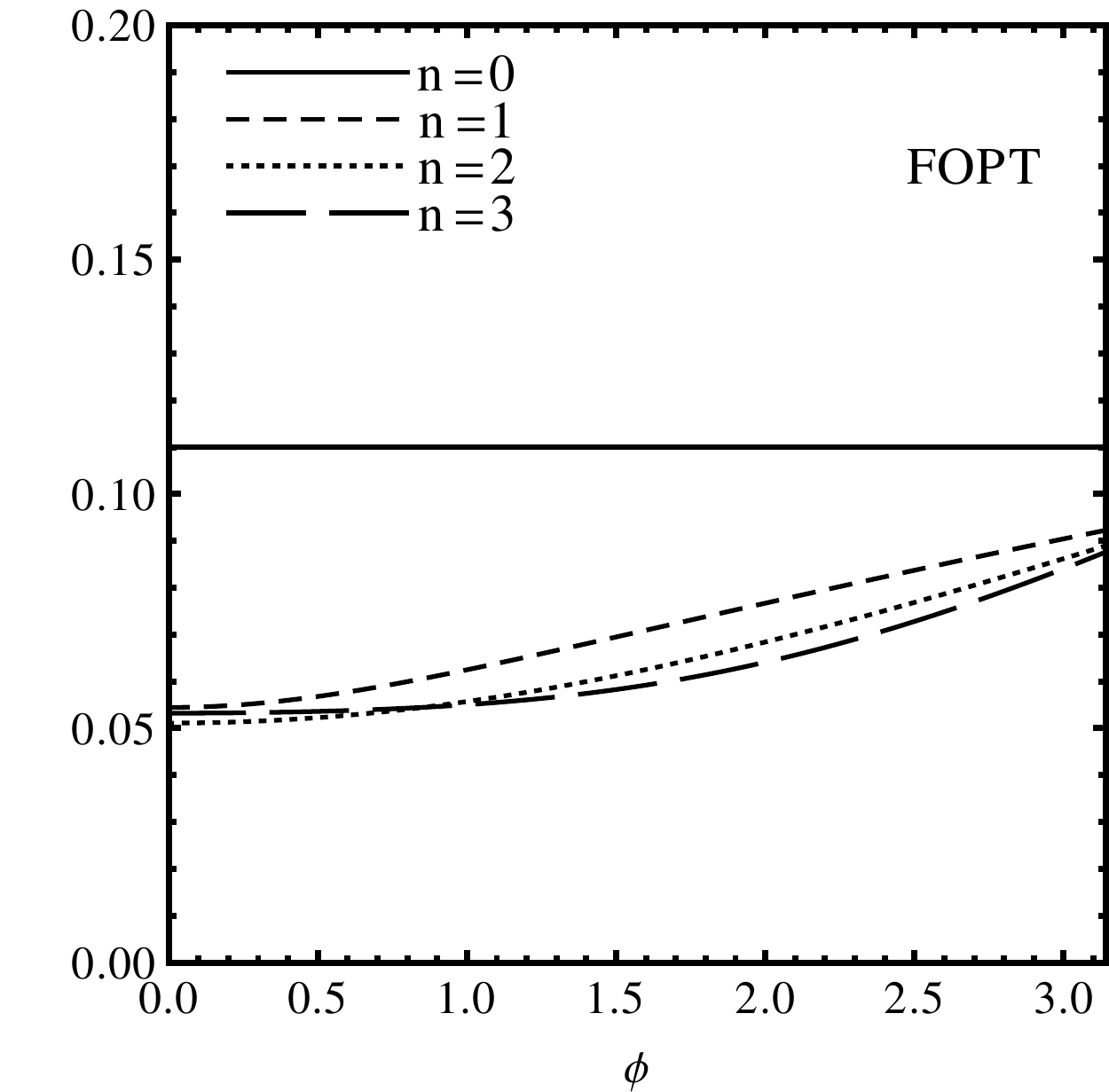}
\caption{Modulus of each successive term of $m_s^2[\mu^2]\Pi^{mq}[\mu^2,Q^2]$ in RGSPT(left) and in FOPT(right) along the circle 
$Q^2 = M_\tau^2 e^{i\phi}$. Here we have taken $m_s(M_\tau^2) = 110$ MeV. \label{fig:conv1}}
\end{center}
\end{figure}

\begin{figure}[ht]
\begin{center}
 \includegraphics[width = 4cm]{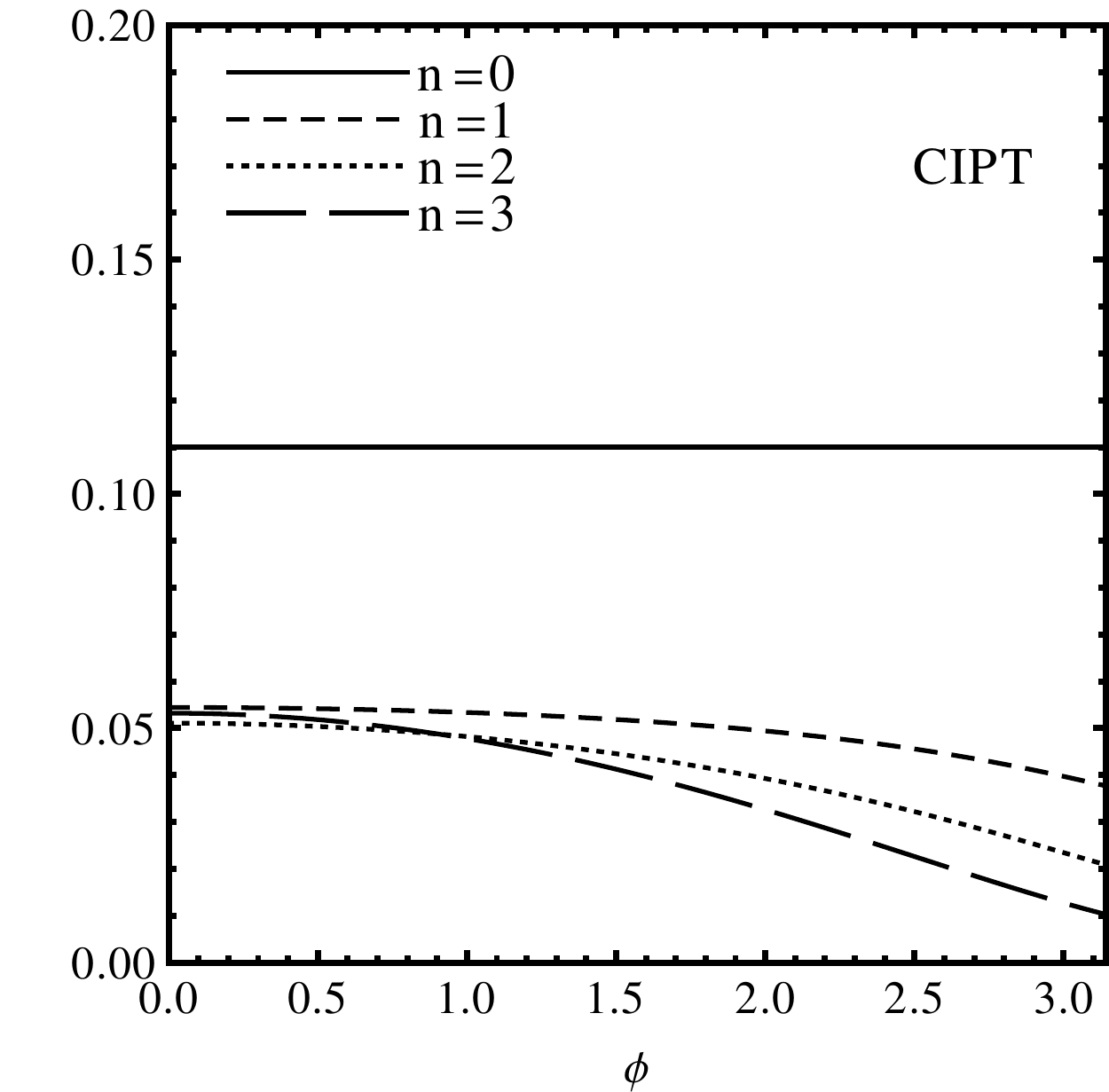}
 \includegraphics[width = 4cm]{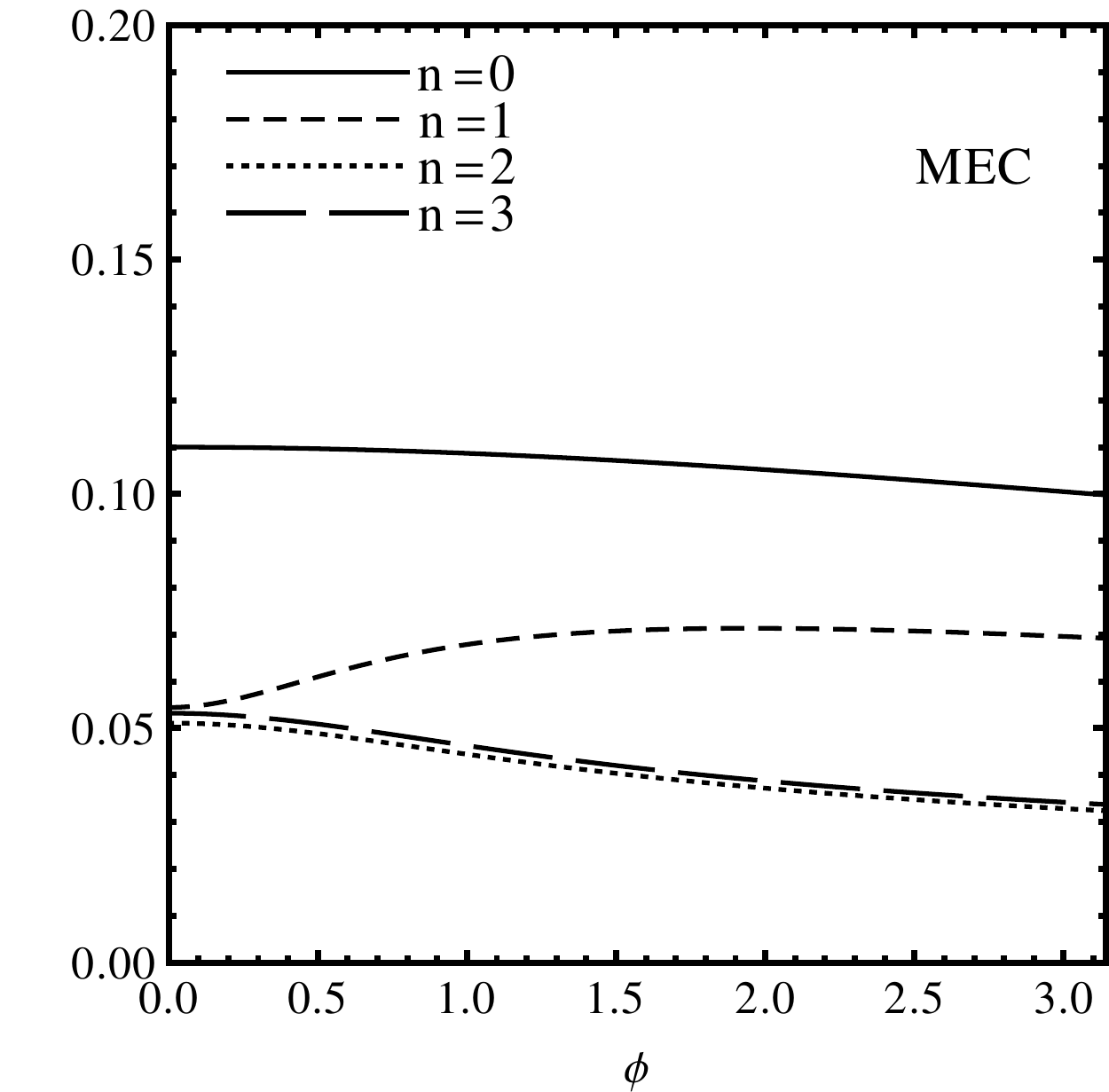}
\caption{Modulus of each successive term of $m_s^2[\mu^2]\Pi^{mq}[\mu^2,Q^2]$ in CIPT(left) and MEC(right) along the circle 
$Q^2 = M_\tau^2 e^{i\phi}$. Here we have taken $m_s(M_\tau^2) = 110$ MeV. \label{fig:conv2} }
\end{center}
\end{figure}

The present work at $\mathcal{O}(\alpha_s^3)$ relies on the use of polarization functions that satisfy homogeneous RG equations
and proceeds through a partial integration equation (\ref{eq:Rkl2}). Before turning to the determinations, we study more closely the new series at this order
and compare with other methods of renormalization group improvements. In Figs.~\ref{fig:conv1} and \ref{fig:conv2} we show the convergence of the 
modulus of $m_s(\mu^2)\Pi^{mq}$ at each successive order in perturbation theory. In the RGSPT and FOPT,
the renormalization scale $\mu^2$ is set at the tau mass scale and therefore the mass parameter does not run along the contour. In CIPT the strong 
coupling constant and the mass are evolved along the contour in the complex plane whereas 
in MEC the evolutions are taken care of by the effective coupling and coefficients of mass parameters. In Fig.~\ref{fig:conv1} we have shown the modulus 
of $m_s(\mu^2)\Pi^{mq}$ in RGSPT and in FOPT. In the case of RGSPT, the higher order terms are smaller and very stable along the circle. 
In FOPT, the higher order terms are small near the 
timelike axis but increase near $\phi=0$. Therefore, FOPT is less stable along the integration contour. In CIPT, shown in Fig.~\ref{fig:conv2},
the overall characteristics of the successive terms are similar to that in RGSPT, but it has better stability along the contour compared to CIPT. 
In MEC, the the next-to-leading order term slightly increases beyond $\phi=0.5$ but remains almost constant beyond that. 
Note that in MEC and CIPT, there is no clear order of the product $m_s^2(\mu^2)\Pi^{mq}(\mu^2)$ and we therefore show the successive order terms
by taking $m_s^2(\mu^2)$ and $\Pi^{mq}(\mu^2)$ at desired orders.

\begin{figure}[ht]
\begin{center}
 \includegraphics[width = 4cm]{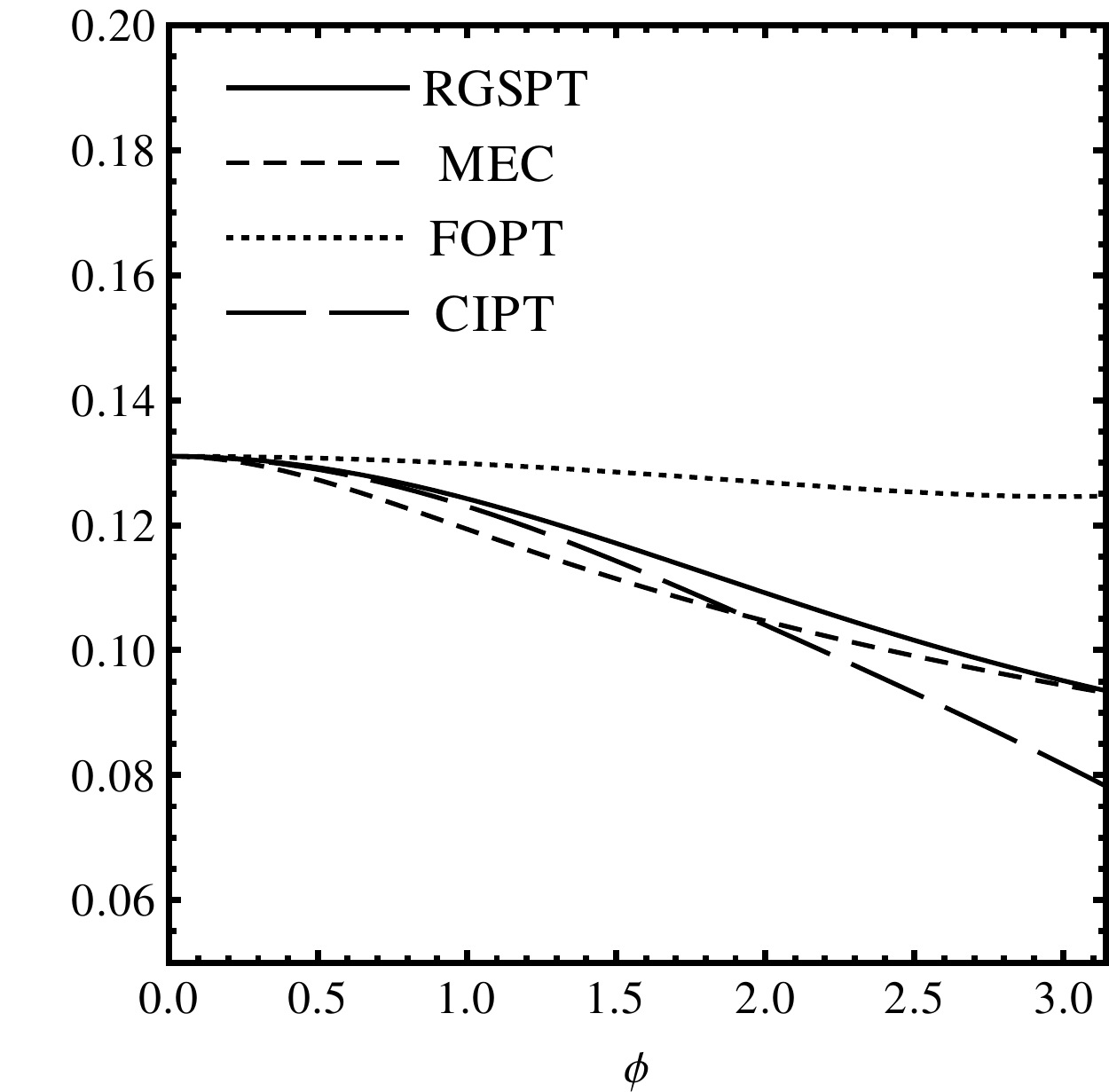}
 \includegraphics[width = 4cm]{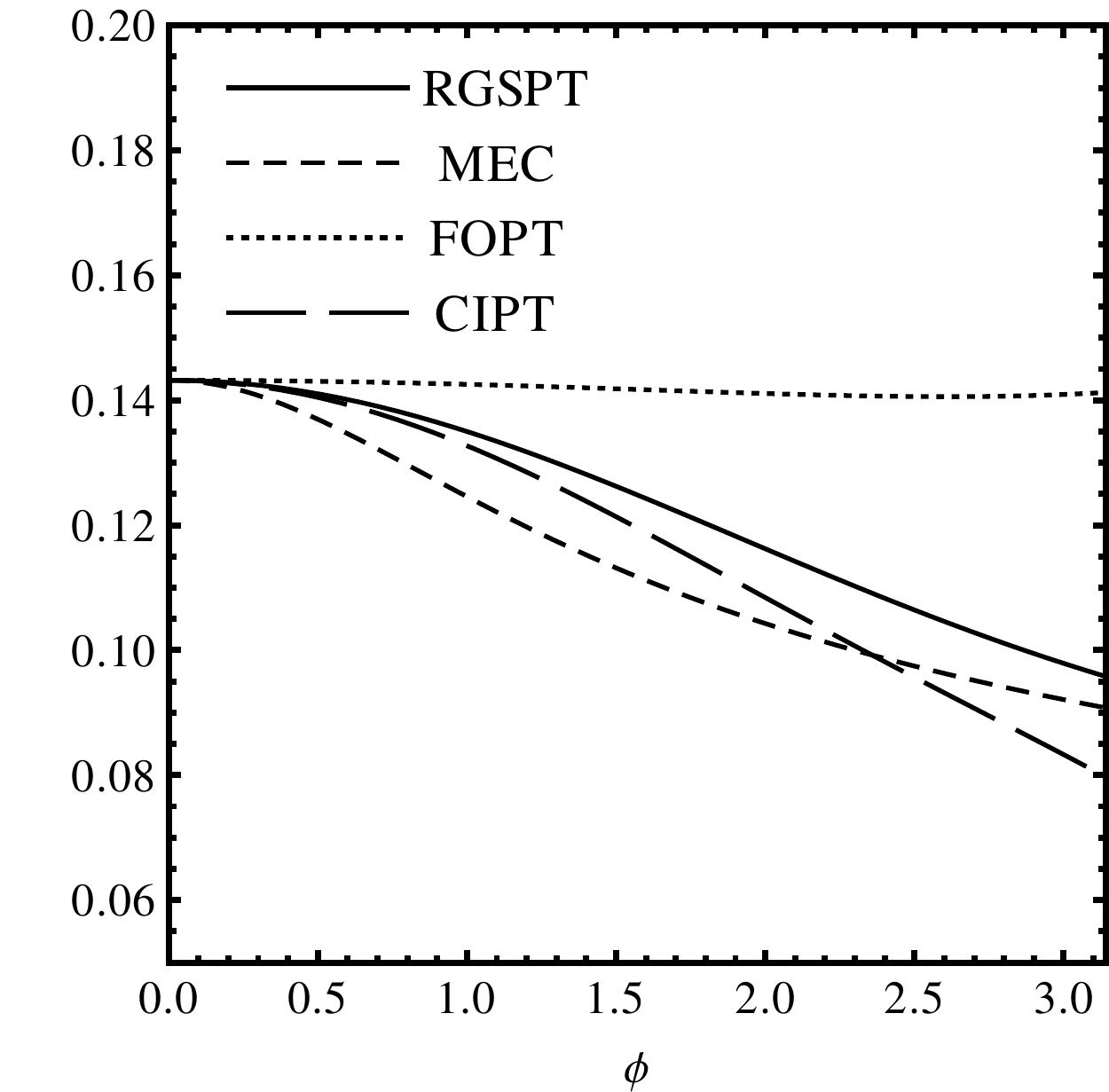}
\caption{Modulus of $m_s^2[\mu^2]\Pi^{mq}[\mu^2,Q^2]$ and $m_s^2[\mu^2] D^{mg}[\mu^2,Q^2]$ along the contour $Q^2 = M_\tau^2 e^{i\phi}$. 
Here the full series is used. Here we have taken $m_s(M_\tau^2) = 110$ MeV.\label{fig:conv}}
\end{center}
\end{figure}

In Fig.~\ref{fig:conv} the convergence of the full series $m_s(\mu^2)\Pi^{mq}$ and $m_s(\mu^2)D^{mg}$ is shown.
Note that RGSPT, CIPT, and MEC have more or less the same convergence behavior. We observe a similar convergence behavior for the function $\Pi^{mq}$.
The different behavior in FOPT is understood from its well-known properties.

\begin{figure}[ht]
\begin{center}
 \includegraphics[height=0.23\textwidth]{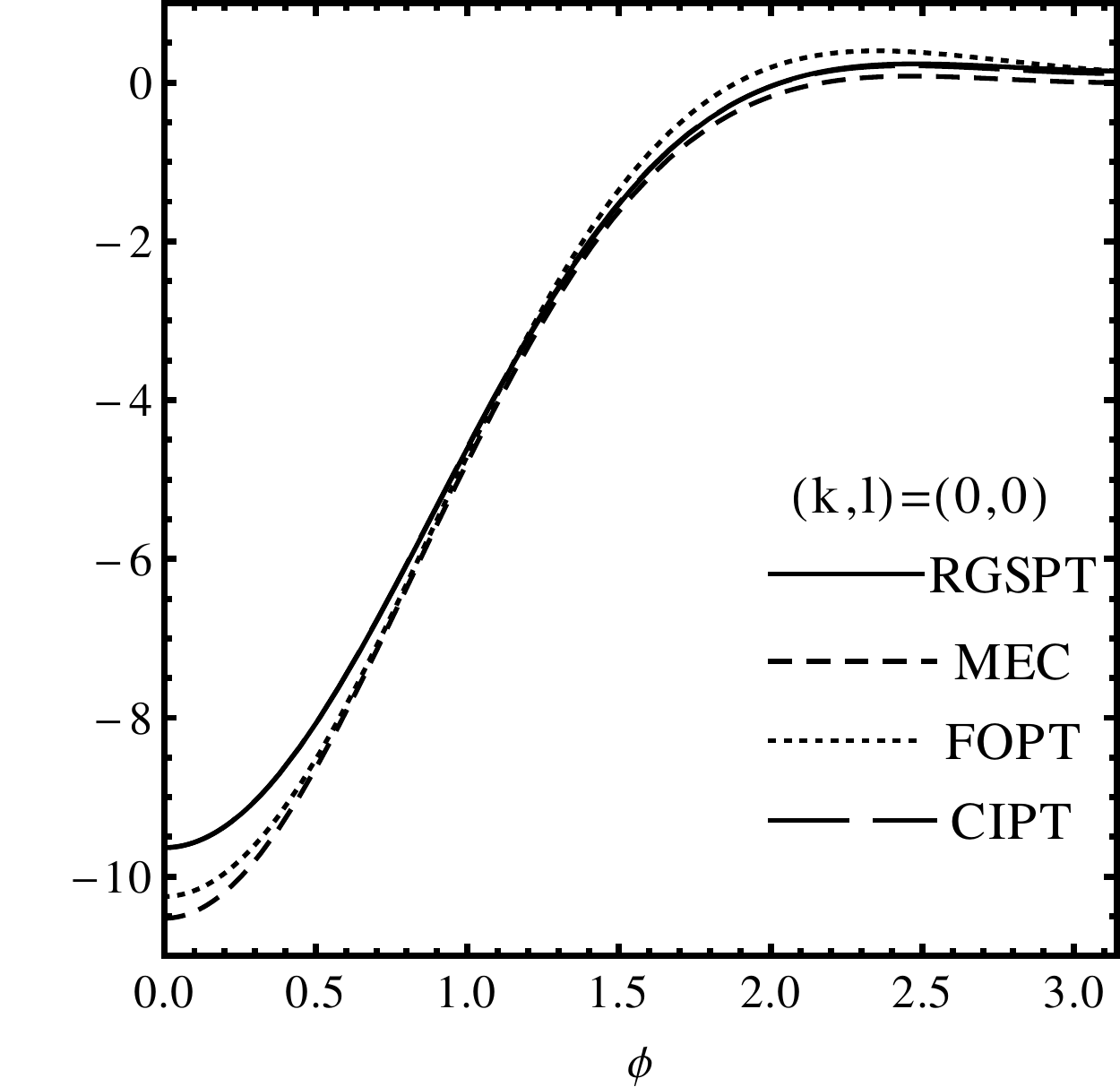}
\includegraphics[height=0.23\textwidth]{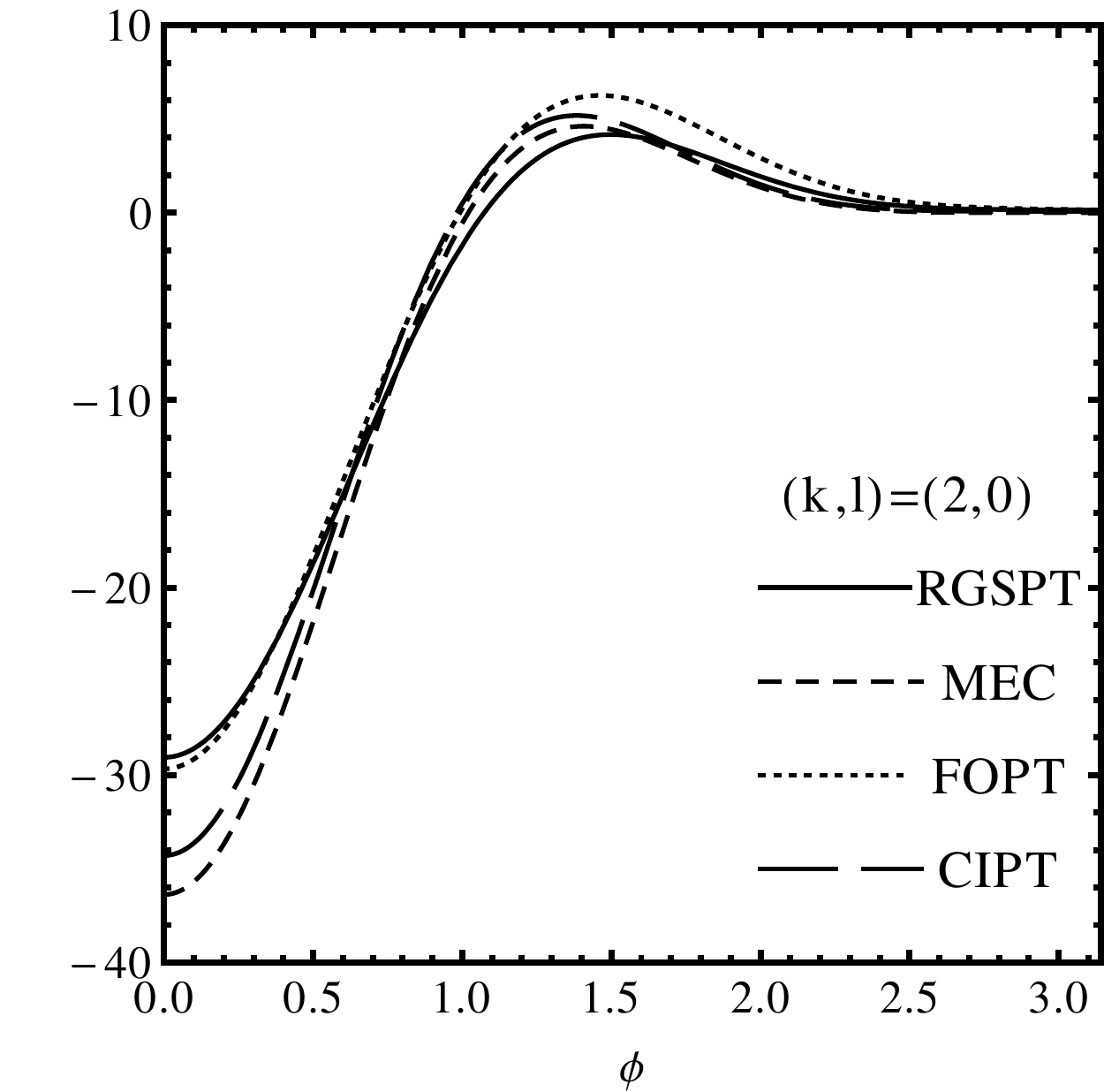}
\caption{The integrand of Eq.~(\ref{eq:Rkl2}) along the contour in the complex plane in RGSPT, CIPT, FOPT, and MEC. \label{fig:mom}}
\end{center}
\end{figure}

In Fig.~\ref{fig:mom} we show the integrand of Eq.~(\ref{eq:Rkl2}) 
along the contour for two different moments (0,0) and (2,0), for all the schemes. This figure shows that the integrand differs significantly 
near $\phi=0$ and $\phi=\pi/2$ between the different schemes.

In Table~\ref{ALEPH} the determinations of strange quark mass from ALEPH data is shown in different renormalization group improvements schemes:
RGSPT, FOPT, CIPT, and MEC. 
Here, the first and the second errors correspond to the experimental and theory uncertainties, respectively.
The determinations in RGSPT at this order can be compared with that
at $\mathcal{O}(\alpha_s^2)$ in Table~\ref{table:msas2} and it shows that the contributions of the $\alpha_s^3$ terms of the polarization functions
are quite significant. In Table~\ref{OPAL} we show our determinations of the strange quark mass from OPAL data for all the schemes.
As can be seen from Tables~\ref{ALEPH} and \ref{OPAL}, the determinations in RGSPT is in agreement with other three schemes. 
The FOPT gives the most stable results whereas the determinations in MEC are smaller than in the other three schemes.
The stability is similar in CIPT and in RGSPT. The remarkable feature of the two tables are that the numbers are consistent with each other
within theory errors along any given row.
These two tables show that the optimal renormalization group improved perturbation series gives reliable determinations of the strange quark mass.
For our final determinations we take the weighted average of different determinations shown in Tables~\ref{ALEPH} and \ref{OPAL}. 
The results at $\tau$ mass scale read   
\begin{equation}\label{eq:final1}
\begin{split}
 m_s(M_\tau^2) = 110.18 \pm 9.67\, {\rm MeV},\quad \text{ALEPH} \, ,\\
 m_s(M_\tau^2) = 76.90 \pm 8.03\, {\rm MeV},\quad \text{OPAL}\, .
\end{split}
\end{equation}
Here the errors are for the smallest error of one individual determination which is (2,0) for ALEPH and (4,0) for OPAL.
We evolve the masses using \cite{Chetyrkin:2000yt,Schmidt:2012az} to 2 GeV and the values read
\begin{equation}\label{eq:final2}
\begin{split}
 m_s(2\text{GeV}) = 106.70 \pm 9.36\, {\rm MeV},\quad \text{ALEPH}\, ,\\
 m_s(2\text{GeV}) = 74.47 \pm 7.77\,  {\rm MeV},\quad \text{OPAL}.
 \end{split}
\end{equation}

\begin{table}[ht]
\renewcommand{\arraystretch}{1.3}
 \begin{tabular}{l c c c c c}\hline \hline
 $(k,l)$  & ~$\widetilde{R}^{kl}_\tau$ from Eq.~(\ref{eq:Rkl}) & ~$\widetilde{R}^{kl}_\tau$ from Eq.~(\ref{eq:Rkl2})  \\\hline
 $(0,0)$   & ~~$-14.86 \pm 0.50$ & $-15.53 \pm 0.50 $  \\                                
  $(1,0)$  &  ~~$-17.79 \pm 0.79$ & $-18.41 \pm 0.78 $  \\
  $(2,0)$  & ~~$-20.76 \pm 1.13$ &  $-21.32 \pm 1.11 $ \\
\hline\hline
 \end{tabular}\caption{ Values of $\widetilde{R}^{kl}_\tau$ at $\mathcal{O}(\alpha_s^2)$ in the RGSPT scheme
 following Eq.~(\ref{eq:Rkl}) (second column) and Eq.~(\ref{eq:Rkl2})(third column). We have 
 defined $R^{kl}_\tau = \widetilde{R}^{kl}_\tau m_s^2(M_\tau^2)/M_\tau^2$.  \label{table:RklThas2}}
 \end{table} 

\begin{table}[ht]
\renewcommand{\arraystretch}{1.3}
 \begin{tabular}{l c c c c c}\hline \hline
 $(k,l)$  & ~$m_s(M_\tau^2)$ from Eq.~(\ref{eq:Rkl})~ & ~$m_s(M_\tau^2)$ from Eq.~(\ref{eq:Rkl2})~ \\\hline
 $(0,0)$   & ~$145.60 \pm 31.66 \pm 4.35 $ & ~$142.69 \pm 30.97 \pm 4.15$ \\                                
  $(1,0)$  &  ~$132.76 \pm 15.41 \pm 5.20 $  & ~$130.80 \pm 15.15 \pm 5.01$ \\
  $(2,0)$  & ~$119.29 \pm 9.69 \pm 5.69 $ & ~$117.97 \pm 9.57 \pm 5.52 $ \\
\hline\hline
 \end{tabular}\caption{ Strange quark mass extracted at $\mathcal{O}(\alpha_s^2)$ from ALEPH data \cite{Barate:1997hv, Schael:2005am, Chen:2001qf}
 in the RGSPT scheme.
 In the second column the moment of the total decay rate $R^{kl}_\tau$ is calculated using Eq.~(\ref{eq:Rkl}) whereas in the third column 
 it is calculated using Eq.~(\ref{eq:Rkl2}). The first and second errors correspond to experimental and theory uncertainties. \label{table:msas2} }
 \end{table} 

\begin{table*}
\centering
\renewcommand{\arraystretch}{1.3}
 \begin{tabular}{l c c c c c}\hline \hline
$(k,l)$ & $\delta R^{kl}_{\tau,\text{expt}}$ & RGSPT & FOPT & CIPT & MEC \\\hline
 $(0,0)$ & ~ $0.374\pm 0.133$  & ~~$131.17 \pm 28.25 \pm 3.98$ & ~~$131.90 \pm 28.42 \pm 4.29$ & ~~$131.09 \pm 28.23 \pm 3.86$ & ~~$126.08 \pm 27.07 \pm 3.39$  \\                                
  $(1,0)$ & ~ $0.398\pm 0.077$ &  ~~$117.71 \pm 13.46 \pm 4.64$ & ~~$125.10 \pm 14.41 \pm 5.63$ & ~~$116.16 \pm 13.26 \pm 4.56$ & ~~$111.43 \pm 12.67 \pm 4.20$ \\
  $(2,0)$  & ~ $0.399\pm 0.053$  & ~~$104.29 \pm 8.31 \pm 4.94$ & ~~$116.48 \pm 9.43 \pm 6.73$ & ~~$101.76 \pm 8.08 \pm 4.88$ & ~~$95.70 \pm 7.54 \pm 4.60$\\
\hline\hline
 \end{tabular}\caption{Determinations of strange quark mass $m_s$ in MeV in different renormalization schemes from the moments of the spectral function 
 $\delta R^{kl}_\tau$ measured by the ALEPH Collaboration \cite{Barate:1997hv, Schael:2005am,Chen:2001qf}. The moments and their experimental values 
 are shown in the first two columns. The third column shows our determinations in RGSPT. The first and the second errors are due to experimental inputs 
 and condensates, respectively. In the rest of the columns we have shown the determinations in FOPT, CIPT, and MEC. \label{ALEPH}}
 \end{table*} 
 
 \begin{table*}
 \centering
\renewcommand{\arraystretch}{1.3}
 \begin{tabular}{l c c c c c}\hline \hline
$(k,l)$ & $\delta R^{kl}_{\tau,\text{expt}}$ & RGSPT & FOPT & CIPT & MEC \\\hline
  $(2,0)$  & ~ $0.264\pm 0.070$  & ~~ $82.14 \pm 13.71 \pm 4.37$ & ~~$91.65 \pm 15.54 \pm 6.49$ & ~~$80.22 \pm 13.34 \pm 4.28$ & ~~$75.60 \pm 12.46 \pm 3.73$ \\
  $(3,0)$ & ~ $0.294\pm 0.055$  & ~~ $78.23 \pm 9.11 \pm 4.60$ & ~~$90.86 \pm 10.90 \pm 7.40$ & ~~$75.66 \pm 8.76 \pm 4.50$ & ~~$69.16 \pm 7.89 \pm 3.86$ \\
  $(4,0)$ & ~ $0.320\pm 0.045$ & ~~ $74.44 \pm 6.49 \pm 4.73$ & ~~$89.96 \pm 8.17 \pm 8.21$ & ~~$71.38 \pm 6.17 \pm 4.63$ & ~~ $62.76 \pm 5.31 \pm 3.92$ \\
\hline\hline
 \end{tabular}\caption{Determinations of strange quark mass $m_s$ in MeV in different renormalization schemes from the moments of the spectral function 
 $\delta R^{kl}_\tau$ measured by the OPAL Collaboration \cite{Abbiendi:2004xa}. The moments and their experimental values are shown in the first two columns. 
 The third column shows our determinations in RGSPT. The first and the second errors are due to experimental inputs and condensates, respectively.
 In the rest of the columns we have shown the determinations in FOPT, CIPT and MEC. \label{OPAL}} 
 \end{table*} 

Our determinations of the strange quark mass can be readily compared with other phenomenological determinations in the literature as well as in lattice QCD.
A good summary of lattice results can be found in \cite{Aoki:2016frl}. The current average of unquenched lattice calculations with $N_f=2, 2+1$ and 2+1+1 yield, 
respectively, $m_s(2{\rm GeV}) = 101(3) {\rm MeV}, 92.0(2.1) {\rm MeV}$ and 93.9(1.1) MeV. Within errors, our determinations are consistent with these
determinations. Based on the experimental results on the $SU(3)$ flavor breaking terms, strange quark mass has been extracted in various phenomenological 
methods in
\cite{Barate:1997hv,Korner:2000wd,Pich:1998yn,Chetyrkin:1998ej,Baikov:2004tk,Chen:2001qf,Kambor:2000dj,Gamiz:2002nu,Gamiz:2004ar,Pich:1999bs,Davier:2001jy}. 
These comparison reveal that barring different phenomenological inputs, in general, our values are in agreement with the past determinations.  
Furthermore, the new RGSPT that we have developed
at this order produces results in striking agreement with those obtained from
FOPT and CIPT also at this order, signaling that the series has converged
very well.

\section{Summary and Conclusion \label{sec:conclusions}}
The renormalization scale dependence of the polarization functions is one of the important sources of uncertainty in the theoretical calculations
of semileptonic $\tau$-decay. Scale dependence can be reduced by resumming the perturbation series which takes into account the 
higher order corrections. In this paper, in the context of the extraction of strange quark mass from $\tau$-decay spectral moments, we apply for the first 
time the method of the optimal renormalization group. This framework makes use of the renormalization group constraints of a perturbation series which results 
in a closed form expressions of all the leading- and next-to-leading order logarithms at each order in perturbation theory. The resummed series exhibits 
reduced scale dependence and we have used them to extract strange quark mass from the measurements of the spectral moments by the ALEPH and the OPAL Collaborations.

With the presently available knowledge of the $\tau$-decay polarization functions, the strange quark mass determination at 
$\mathcal{O}(\alpha_s^2)$ relies on two methods of performing
the contour integration. The first method makes use of the renormalization
group noninvariant function $\Pi^{mg}$ and proceeds through direct integration along the contour; see Eq.~(\ref{eq:Rkl}). 
The second method is based on the partial integration equation (\ref{eq:Rkl2}) and replaces 
the renormalization group noninvariant function by an Adler function
$D^{mg}$ that satisfies the homogeneous RG equation (\ref{eq:RGD}). We have extracted $m_s$ at this order from the ALEPH data using both these methods. Our results show that within
the theory errors, the two methods are consistent with each other. The results are shown in Tables~\ref{table:RklThas2} and \ref{table:msas2}.
One may wish to test the consistency also at $\mathcal{O}(\alpha_s^3)$ provided the relevant coefficients are made available.

At $\mathcal{O}(\alpha_s^3)$ only the second method of integration is applicable. At this order we have studied in detail the convergence
of the RGSPT series along the integration contour. The results are shown in Figs.~\ref{fig:conv1}, \ref{fig:conv2}, and \ref{fig:conv}. 
Here we have compared RGSPT with several other schemes of RG improvements such as FOPT, CIPT, and MEC.
We find that the series converges very well and are comparable to each other.  
In Fig.~\ref{fig:mom} we have done a comparative study of the behavior of the integrand of Eq.~(\ref{eq:Rkl2}) along the contour in these schemes.
Our determinations of the strange quark mass using ALEPH and OPAL data are shown in Tables~\ref{ALEPH} and \ref{OPAL}, respectively.
These determinations are the most up to date with latest theoretical and experimental inputs.
Note that, at present, the determinations are dominated by the experimental errors.
Note that the strange quark mass determination based on contour-improved perturbation theory in an
effective scheme was undertaken in Ref.~\cite{Korner:2000wd}. In this paper we have updated the determinations to $\alpha_s^3$
with the latest inputs from the ALEPH and OPAL Collaborations and extended these to several moments with nonzero $k$.

By comparing the strange quark mass determined in different schemes, the important picture
that emerges from our study is that, in general, the extracted mass is
insensitive to the choice of schemes of renormalization group improvements.
As a result we could argue that in the future when the data improves, excellent determinations of $m_s$ could be made.
The fact that all schemes giving similar numbers implies that we have good convergence at this order
and so it is desirable to have the results without partial integration.

For our final determinations, we have taken the weighted average of individual determinations made with ALEPH and OPAL
data. These are shown in Eq.~(\ref{eq:final1}). For easy comparison with other methods, these values are evolved to
2 GeV in Eq.~(\ref{eq:final2}).

Although optimal renormalization has been around for over fifteen years, it has not been
developed for the extraction of the $s$-quark mass.  Our work has filled this gap in the
treatment of this quantity. 
We have relied on experimental information that has been
the basis of such extractions in the past, and we get encouraging results.  While a
full analysis of polarization functions may in the future lead to more stable determinations,
the full framework has been provided in this work.

\section*{Acknowledgements}
It is a great pleasure to thank Irinel Caprini for suggestions and improvements on the manuscript. 
B. A. thanks Daniel Wyler and the University of Zurich for their hospitality when part of this work was done.
D. D. thanks Kim Maltman, Konstantin Chetyrkin, and Diogo Boito for helpful correspondence.
B. A. is partly supported by the MSIL Chair of the Division of Physical and Mathematical Sciences, Indian Institute
of Science.

\appendix
\section{Perturbative coefficients of $\beta$ and $\gamma$ functions \label{app:betagamma}}
Our definitions of the QCD $\beta$ function and the anomalous dimension matrix $\gamma_m$ are given in Eqs.~(\ref{def:beta}) and (\ref{def:gamma}),
respectively. Beyond the leading order these equations cannot be solved exactly. One can, however,
obtain perturbative solutions given that their values are known for some starting scale $\mu_0$, $a_s(\mu_0)$, $\mu(\mu_0)$.
The perturbative solutions read \cite{Chetyrkin:1997wm}
\begin{align}
 &a_s(\mu^2, Q^2)  = a_s + a_s^2 \beta_0 L + a_s^3 (\beta_1 L + \beta_0^2 L ) + a_s^4 \Big(\beta_2 L^2 \, \nn\\
 &+ \frac{5}{2}\beta_0 \beta_1 + \beta_0^3  L^3 \Big) + a_s^5 \Big(\beta_3 L + \frac{3}{2} \beta_1^2 L^2  \, \nn\\
 &+ 3\beta_0 \beta_2 L^2 + \frac{13}{3} \beta_0^2 \beta_1 L^3  + \beta_0^4 L^4 \Big)\, ,\\
 &m_s(\mu^2,Q^2) =  m_s\Bigg[1 + a_s \gamma_0 L + a_s^2 \Big( \frac{\gamma_0}{2}(\beta_0+\gamma_0) L^2  \, \nn\\
 &+ \gamma_1 L \Big)  + a_s^3 \Bigg( \gamma_0\Big( \frac{\beta_0^2}{3} + \frac{\beta_0\gamma_0}{2} + \frac{\gamma_0^2}{6} \Big) L^3 + \Big(\beta_0\gamma_1 \, \nn\\
 & + \frac{\beta_1\gamma_0}{2} + \gamma_0\gamma_1 \Big) L^2 - \gamma_2 L \Bigg)
 + a_s^4 \Bigg( \frac{1}{4}  L^4 \Big(\beta _0+\frac{\gamma _0}{2}\Big) \, \nn\\
&\times \Big(\beta _0+\frac{\gamma _0}{3}\Big) \gamma _0 \Big(\beta _0+\gamma _0 \Big)+ L^3 \Big(\frac{1}{2} \beta _1 \gamma _0
   \Big(\frac{5 \beta _0}{3}+\gamma _0\Big) \,\nn\\ 
   &+\Big(\beta _0+\frac{\gamma _0}{2}\Big) (\beta _0+\gamma _0) \gamma _1\Big)+L^2
   \Big(\frac{\beta _2 \gamma _0}{2}+(\beta _1+\frac{\gamma _1}{2}) \gamma _1\,\nn\\&+(\frac{3 \beta _0}{2}+\gamma _0) \gamma _2\Big)+ L
   \gamma _3 \Bigg)  \Bigg]\, 
\end{align}
In both these expressions, in the right-hand side, $a_s$ and $m_s$ are at the scale $\mu_0$, $a_s=a_s(\mu_0)$ and $m_s=m_s(\mu_0)$.
The coefficients of the $\beta$ function for $n_f$ active quark flavors are
\begin{align}
\beta_0 &= \frac{11}{4}-\frac{1}{6}n_f\, ,\nn\\
\beta_1 &= \frac{51}{8} - \frac{19}{24}n_f\, ,\nn\\
\beta_2 &= \frac{2857}{128} - \frac{5033}{1152} n_f + \frac{325}{3456}n_f^2\, \\
\beta_3 & = \frac{149753}{1536} - \frac{1078361}{41472} n_f + \frac{50065}{41472} n_f^2 + \frac{1093}{186624} n_f^3  \, \nn\\
        & + \frac{891}{64} \zeta(3) - \frac{1627}{1728} n_f \zeta(3)  + \frac{809}{2592} n_f^2 \zeta(3)\, .\nn
\end{align}
The coefficients of $\gamma_m$ functions for $n_f$ active flavors read
\begin{align}
\gamma_0 &= 1\, ,\nn\\
\gamma_1 &= \frac{101}{24} - \frac{5}{36}n_f\, ,\nn\\
\gamma_2 &= \frac{1249}{64} - \frac{277}{216} n_f - \frac{35}{1296} n_f^2 - \frac{5}{6}n_f \zeta(3)\, ,\\
\gamma_3 &= \frac{4603055}{41472} - \frac{91723}{6912} n_f + \frac{2621}{31104}n_f^2 - \frac{83}{15552}n_f^3 \nn\\
         &   + \frac{11\pi^4}{288}n_f - \frac{\pi^4}{432}n_f^2 + \frac{530}{27}\zeta(3) - \frac{2137}{144}\zeta(3)n_f \nn\\
         &   + \frac{25}{72}\zeta(3)n_f^2 + \frac{1}{108}\zeta(3)n_f^3 - \frac{275}{8}\zeta(5) + \frac{575}{72}\zeta(5)n_f\, .\nn
\end{align}
Here $\zeta(n)$ are the Reimann $\zeta$ functions.
\section{Perturbative expansion of $\Pi^{mg,mq}$ and $D^{mg}$ \label{app:spectral}}
The polarization function $\Pi^{mg}[\mu^2,Q^2]$ has been calculated to order $\alpha_s^2$ in perturbation theory 
\cite{Gorishnii:1986pz,Surguladze:1994bx,Chetyrkin:1996hm} and it reads
\begin{align}\label{eq:PimgUnsummed}
& \Pi^{mg}[\mu^2,Q^2] = \ln\frac{\mu^2}{Q^2} + a_s\Big( \frac{25}{4} - 4\zeta(3) + \frac{5}{3}\ln\frac{\mu^2}{Q^2}  \, \nn\\
& + \ln^2\frac{\mu^2}{Q^2} \Big) + a_s^2 \Bigg( \frac{18841}{432} - \frac{3607}{54}\zeta(3) + \frac{1265}{27}\zeta(5)  \,\nn\\
&- \frac{\pi^4}{360} + \frac{4591}{144}\ln\frac{\mu^2}{Q^2} - \frac{35}{2}\ln\frac{\mu^2}{Q^2}\zeta(3) + \frac{22}{3}\ln^2\frac{\mu^2}{Q^2} \, \nn\\
&+ \frac{17}{12}\ln^3\frac{\mu^2}{Q^2} \Bigg) \, .
\end{align}
The Adler function $D^{mg}$ defined in Eq.~(\ref{eq:defAdler}) is known to order $\alpha_s^3$ in perturbation theory. The expression
can be found in Ref.\cite{Baikov:2002uw}, which defined it at normalization scale $\mu^2=Q^2$. The logarithmic terms have been
generated using the renormalization group, Eq.~(\ref{eq:RGD}). The expression reads
\begin{align}\label{eq:DmgUnsummed}
&D^{mg}[\mu^2,Q^2] = 1 + a_s\Bigg(\frac{5}{3} + 2\ln\frac{\mu^2}{Q^2}\Bigg) + a_s^2 \Bigg( \frac{4591}{144}  \,\nn\\
& - \frac{35}{2}\zeta(3) +\frac{44}{3}\ln\frac{\mu^2}{Q^2} + \frac{17}{4}\ln^2\frac{\mu^2}{Q^2} \Bigg) + a_s^3 \Bigg( \frac{1967833}{5184}  \,\nn\\
&- \frac{\pi^4}{36} -\frac{11795}{24}\zeta(3) + \frac{33475}{108}\zeta(5) + \frac{4633}{18}\ln\frac{\mu^2}{Q^2}  \,\nn\\
&-\frac{475}{4}\ln\frac{\mu^2}{Q^2}\zeta(3) + \frac{237}{4}\ln^2\frac{\mu^2}{Q^2} +\frac{221}{24}\ln^3\frac{\mu^2}{Q^2} \Bigg)\, .
\end{align}
Finally, we give the expression of $\Pi^{mq}$, which is known to $\alpha_s^3$ in perturbation theory \cite{Korner:1999kw,Baikov:2004tk}.
This function was also defined at the renormalization scale $\mu^2=Q^2$ and the logarithmic terms have been generated using RG
equation (\ref{eq:RGPimq}). The expression reads
\begin{align}\label{eq:PimqUnsummed}
& \Pi^{mq}[\mu^2,Q^2] = 1 + a_s\Bigg(\frac{7}{3}+2\ln\frac{\mu^2}{Q^2} \Bigg) + a_s^2 \Bigg( \frac{13981}{432}\, \nn\\ 
&+ \frac{323}{54}\zeta(3) - \frac{520}{27}\zeta(5) + \frac{35}{2}\ln\frac{\mu^2}{Q^2} + \frac{17}{4}\ln^2\frac{\mu^2}{Q^2} \Big) \, \nn\\ 
&+ a_s^3 \Bigg( \frac{2092745}{5184} - \frac{\pi^4}{36} + \frac{14713}{648} \zeta(3) + \frac{61}{2} \zeta(3)^2 \,\nn\\ 
& - \frac{41065}{108} \zeta(5) + \frac{79835}{648}\zeta(7) + \frac{14485}{54}\ln\frac{\mu^2}{Q^2}  \,\nn\\ 
& - \frac{3380}{27}\zeta(5)\ln\frac{\mu^2}{Q^2} + \frac{3659}{108}\zeta(3)\ln\frac{\mu^2}{Q^2} + \frac{1643}{24}\ln^2\frac{\mu^2}{Q^2} \,\nn\\  
& + \frac{221}{24}\ln^3\frac{\mu^2}{Q^2} \Bigg)\, ,
\end{align}
In these expressions the QCD coupling constant is taken at generic value of the t'Hooft mass scale $\mu$ and $n_f=3$ is the number of active 
quark flavors. All the functions are
normalized in the $\overline{{\rm MS}}$ scheme. 

\section{Coefficients of closed form summations \label{app:closed}}
The closed form expressions of the intermediate quantities $\Pi^{mg}_j[a_s, L]$, $D^{mg}_j[a_s,L]$, and $\Pi^{mq}_j[a_s,L]$
are given in Eqs.~(\ref{eq:sol0})-(\ref{eq:sol3}). The coefficients $B$, $C$, $D$, $E$, $F$, $G$, $H$, $K$, $M$, $N$, $P$, $Q$, $R$, $S$, $T$, $U$, $V$, $Y$,
and $T_{1\cdots 9}$ that appear in these equations can be written in terms of the leading coefficients of the perturbation
series at each order and the coefficients of $\beta$ and the $\gamma$ functions
 \begin{align}
& B = \frac{c_{0,0} (A\beta_1-2\gamma_1) }{\beta_0}\, ,\quad C = -A\frac{c_{0,0}\beta_1}{\beta_0}\, ,\\
& D = -B\frac{\beta_1}{\beta_0} + AB\frac{\beta_1}{\beta_0} + A\frac{c_{0,0}\beta_2 }{\beta_0} - 2B \frac{\gamma_1}{\beta_0} - 2\frac{c_{0,0}\gamma_2}{\beta_0}\, ,\\
& E = -\frac{2AB\beta_1 + C \beta_1 - A c_{1,0}\beta_1 + A c_{0,0}\beta_2 - 2B\gamma_1 + 2 c_{1,0}\gamma_1 }{\beta_0}\, ,\\
&  F = \frac{C(A\beta_1-2\gamma_1)}{\beta_0}\, ,\quad G = \frac{( B + AB +C - (1+A)c_{1,0} )\beta_1 }{\beta_0}\, ,\quad\\
& H = -\frac{(1+A)C\beta_1}{\beta_0} \, ,\quad
K = -D\frac{\beta_1}{\beta_0} + AD\frac{\beta_1}{2\beta_0} - B\frac{\beta_2}{\beta_0} \,\nn\\
&+ AB\frac{\beta_2}{\beta_0} + A\frac{c_{0,0}\beta_3}{\beta_0} - D\frac{\gamma_1}{\beta_0} - 2B\frac{\gamma_2}{\beta_0} - 2B\frac{\gamma_3}{\beta_0}\, ,
      \end{align}

\begin{widetext}
\begin{align}
 & M = -\frac{A D {\beta_1}}{2 \beta_0 }-\frac{E \beta_1}{\beta_0}+\frac{A E \beta_1}{\beta_0}-\frac{A F \beta_1}{\beta_0}-
      \frac{2 A B \beta_2}{\beta_0}-\frac{C\beta_2}{\beta_0}+\frac{A c_{1,0} \beta_2}{\beta_0}-\frac{A c_{0,0} \beta_3}{\beta_0}
      -\frac{2 E \gamma_1}{\beta_0}+\frac{2 F \gamma_1}{\beta_0}+\frac{2 B \gamma_2}{\beta_0}\, \nn\\
      &\quad\quad -\frac{2c_{1,0}\gamma_2}{\beta_0} +\frac{2 B \gamma_3}{\beta_0}-\frac{2 c_{1,0}\gamma_3}{\beta0}\, ,\\
& N = -(F-AF)\frac{\beta_1}{\beta_0} + AC\frac{\beta_2}{\beta_0} - 2F \frac{\gamma_1}{\beta_0} - 2C\frac{\gamma_2}{\beta_0} - 2C\frac{\gamma_3}{\beta_0}\, ,\\
& P = -\frac{A D \beta _1}{2 \beta _0}-\frac{E \beta_1}{\beta_0}-\frac{2 A E \beta_1}{\beta_0}+\frac{2 F \beta_1}{\beta_0}
    +\frac{2 A F \beta_1}{\beta_0}-\frac{G \beta_1}{\beta_0}+\frac{A c_{2,0} \beta_1}{\beta_0}+\frac{B \beta_2}{\beta_0}+\frac{AB\beta_2}{\beta_0}
    +\frac{C \beta_2}{\beta_0}-\frac{c_{1,0} \beta_2}{\beta_0}\, \nn\\ 
    &\quad\quad-\frac{A c_{1,0} \beta_2}{\beta_0}+\frac{D \gamma_1}{\beta_0}
    +\frac{2 E \gamma_1}{\beta_0}-\frac{2 F \gamma_1}{\beta_0}-\frac{2 c_{2,0} \gamma_1}{\beta_0}\, ,\\
&Q = -\frac{F \beta _1}{\beta _0}-\frac{A F \beta _1}{\beta _0}+\frac{A G \beta _1}{\beta _0}-\frac{H \beta _1}{\beta _0}-\frac{C \beta _2}{\beta _0}-\frac{A
   C \beta _2}{\beta _0}-\frac{2 G \gamma _1}{\beta _0}\, ,\\
   & R = AH\frac{\beta_1}{2\beta_0} - H\frac{\gamma_1}{\beta_0}\, ,\quad
U = \frac{D \beta _1}{\beta _0}+\frac{A D \beta _1}{2 \beta _0}+\frac{2 E \beta _1}{\beta _0}+\frac{A E \beta _1}{\beta _0}-\frac{2 F \beta _1}{\beta
   _0}-\frac{A F \beta _1}{\beta _0}+\frac{G \beta _1}{\beta _0}-\frac{2 c_{2,0} \beta _1}{\beta _0}-\frac{A c_{2,0} \beta _1}{\beta _0} \, ,\\
& V = -2G\frac{\beta_1}{\beta_0} - AG\frac{\beta_1}{\beta_0} + H\frac{\beta_1}{\beta_0}\, ,\quad Y = -(2+A)\frac{H\beta_1}{2\beta_0}\, .
\end{align}
For the expressions of $\Pi^{mg}_j$, $D^{mg}_j$, and $\Pi^{mq}_j$, the coefficients $c_{n,0}$ correspond to the coefficients 
$c^{mg}_{n,0}, d^{mg}_{n,0}$, and $c^{mq}_{n,0}$, respectively. The coefficients $T_1$--$T_9$ read
\begin{align}
& T_1 = \frac{-t_{2,0}\gamma_0 + t_{1,0}\gamma_1 }{\beta_0\gamma_0}\, ,\quad T_2 = \frac{t_{1,0}(\beta_1\gamma_0-\beta_0\gamma_1)}{\beta_0^2\gamma_0}\, ,\quad 
  T_3 = -\frac{t_{1,0}\beta_1}{\beta_0^2}\, ,\\
& T_4 = \frac{-(1+A) t_{30} \gamma _0-T_1 \gamma _0 \left(\beta _1+2 \gamma _1\right)+(1+A) t_{10} \gamma _2}{(1+A) \beta _0 \gamma _0}\, ,\\
& T_5 = -\frac{T_2 \beta _1}{\beta _0}+\frac{A T_2 \beta _1}{\beta _0}+\frac{A t_{10} \beta _2}{2 \beta _0 \gamma _0}-\frac{2 T_2 \gamma _1}{\beta_0}
         -\frac{t_{10} \gamma _2}{\beta _0 \gamma _0}\, ,\\
& T_6 = -\frac{2 T_3 \beta _1 \gamma _0+A^2 \left(t_{10} \beta _2+4 T_2 \beta _1 \gamma _0\right)-4 \left(T_1+T_2\right) \gamma _0 \gamma _1+A \left(t_{10} \beta
   _2+2 \left(T_1+2 T_2+T_3\right) \beta _1 \gamma _0-4 T_2 \gamma _0 \gamma _1\right)}{2 (1+A) \beta _0 \gamma _0}\, ,\\
& T_7 = \frac{T_3 \left(A \beta _1-2 \gamma _1\right)}{\beta _0}\, ,\quad T_8 = \frac{\left(T_1+T_2+A T_2+T_3\right) \beta _1}{\beta _0} \, ,\quad
T_9 = -\frac{(1+A) T_3 \beta _1}{\beta _0}\, .
\end{align}
The coefficients  $c^{mg}_{n,0}, d^{mg}_{n,0}$, and $c^{mq}_{n,0}$ can be obtained from Eqs.~(\ref{eq:PimgUnsummed}), (\ref{eq:DmgUnsummed}), and (\ref{eq:PimqUnsummed})
following the definitions equation (\ref{eq:PimqPert}). The coefficients $t_{n,0}$ can be obtained in the similar way from the expression of $\gamma_m^{VV}$ given in 
\cite{Chetyrkin:1996hm,Chetyrkin:1996sr} which we have defined as $\gamma_m^{VV} = 6\sum_{n=0}^\infty \sum_{k=0}^n t_{n,k} a_s^n L^k$. These are given below
for completeness
\begin{eqnarray}
c^{mg}_{0,0} &=& 0\, ,\quad c^{mg}_{1,0} = \frac{25}{4} - 4\zeta(3)\, ,\quad c^{mg}_{2,0} = \frac{18841}{432} - \frac{\pi^4}{360} - \frac{3607}{54}\zeta(3)+\frac{1265}{27}\zeta(5)\, ,\\
 d^{mg}_{0,0} &=& 1 ,\quad d^{mg}_{1,0} = \frac{5}{3}\, ,\quad d^{mg}_{2,0} = \frac{4591}{144} - \frac{35}{2}\zeta(3)\, ,\quad
 d^{mg}_{3,0} = \frac{1967833}{5184}-\frac{\pi^4}{36}-\frac{11795}{24}\zeta(3) + \frac{33475}{108}\zeta(5)\, ,\\
 c^{mq}_{0,0} &=& 1\, ,\quad c^{mq}_{1,0}=\frac{7}{3}\, ,\quad c^{mq}_{2,0} = \frac{13981}{432}+\frac{323}{54}\zeta(3)-\frac{520}{27}\zeta(5)\, ,\\
 c^{mq}_{3,0} &=& \frac{2092745}{5184}-\frac{\pi^4}{36}+\frac{14713}{648}\zeta(3)+\frac{61}{2}\zeta(3)^2 - \frac{41065}{108}\zeta(5) + \frac{79835}{648}\zeta(7)\, ,\\
 t_{1,0} &=& \frac{5}{3}\, ,\quad t_{2,0} = \frac{455}{72} - \frac{n_f}{3} - \frac{1}{2}\zeta(3)\, ,\\
 t_{3,0} &=& \frac{157697}{5184} - \frac{14131}{7776}n_f - \frac{1625}{11664}n_f^2 + \frac{\pi^4}{48} - \frac{11\pi^4}{1080}n_f - \frac{1645}{216}\zeta(3)
          - \frac{13}{9}n_f\zeta(3) + \frac{1}{9} n_f^2\zeta(3) + \frac{65}{12}\zeta(5)\, .
\end{eqnarray}
\end{widetext}

\end{document}